# Ultrafast control of spin order by linearly polarized light in noncollinear antiferromagnetic metals

J. Kimák[1], M. Nerodilová[1], K. Carva[1], S. Ghosh[2], J. Železný[2], T. Ostatnický[1], J. Zemen[3],

F. Johnson[4], D. Boldrin[5], F. Rendell-Bhatti[5], B. Zou[6], A.P. Mihai[6], X. Sun[7], F. Yu[7],

E. Schmoranzerová[1], L. Nádvorník[1], L.F. Cohen[8], and P. Němec[1,*]

[1]Faculty of Mathematics and Physics, Charles University, Ke Karlovu 3, 121 16 Prague 2, Czech Republic
[2]Institute of Physics ASCR, v.v.i., Cukrovarnická 10, 162 53 Prague 6, Czech Republic
[3]Faculty of Electrical Engineering, Czech Technical University, Technická 2, Prague 6, 166 00, Czech Republic
[4]Cavendish Laboratory, University of Cambridge, JJ Thomson Ave, Cambridge, CB3 0HE, United Kingdom
[5]SUPA, School of Physics and Astronomy, University of Glasgow, Glasgow, G12 8QQ, United Kingdom
[6]LoMaRe Technologies Ltd., 6 London Street, EC3R 7LP London, United Kingdom
[7]LoMaRe Chip Technology Changzhou Co., Ltd., 400 Hanjiang Road, Changzhou, China
[8]Department of Physics, Blackett Laboratory, Imperial College London, London, SW7 2AZ, United Kingdom

**The non-thermal optical control of magnetic order offers a promising route to ultrafast, energy-efficient information technologies. Although optical manipulation of magnetism in metals has been extensively studied, experimentally demonstrated effects have so far been limited to heat-driven dynamics or helicity-dependent mechanisms. Here, we report ultrafast non-thermal control of spin order in noncollinear antiferromagnetic Mn-based antiperovskite nitrides $Mn_3NiN$ and $Mn_3GaN$, driven solely by the polarization orientation of linearly polarized femtosecond laser pulses. Using time-resolved magneto-optical pump–probe experiments based on the Voigt effect, we observe sub-picosecond changes in magnetic order followed by picosecond relaxation. The magneto-optical response depends on the relative orientation of the pump and probe polarization planes, with linear-polarization dependence reaching up to 95%, a value unprecedented in metallic magnets. This phenomenon is observed in two different materials and persists over a wide range of excitation wavelengths, fluences, and temperatures, demonstrating its robustness. Symmetry analysis and microscopic modeling indicate that optically induced torques alone cannot fully explain the observed dynamics. We therefore propose laser-induced formation of transient spin-spiral states as a possible excitation mechanism.**

---

[*] Electronic mail: petr.nemec@matfyz.cuni.cz

Ultrafast control of magnetic order using femtosecond laser pulses has attracted attention since its discovery in 1996, when demagnetization of a thin Ni film in less than 1 ps was demonstrated[1]. These experiments enable the study of magnetic system dynamics on a time scale corresponding to the exchange interaction (tens or hundreds of femtoseconds), which is much shorter than the time scales of spin-orbit interaction (picoseconds) and magnetization precession (hundreds or thousands of picoseconds in ferromagnets)[2]. At this ultrafast time scale, excitation by a femtosecond laser pulse places a magnetic medium in a highly nonequilibrium state where the conventional macrospin approximation fails and the different reservoirs of a magnetic system, such as the magnetically ordered spins, the electron system and the lattice, are dynamically isolated[2-4]. Additionally, ultrafast excitation by laser pulses can provide access to various chiral spin textures (e.g. skyrmions) that otherwise remain hidden in magnetic materials during adiabatic field cycling[5-7].

Laser pulses can interact with a magnetic system through many mechanisms. The simplest are thermal effects, where the absorption of light leads to a transfer of energy from the laser pulse to the material and consequently to an increase in spin temperature. Since the direct pumping of energy from photons to spins is not effective, the light pumps the energy through the electron and phonon subsystems[2-4]. The second group are non-thermal photomagnetic effects, in which the absorption of light by certain electronic states leads to a direct change in magnetic parameters, such as magnetocrystalline anisotropy or exchange interaction[2]. The third group are non-thermal optomagnetic effects, which include the inverse Faraday and Cotton-Mouton effects[8,9], that are not related to light absorption but are due to stimulated Raman scattering[2].

The identification of a precise excitation mechanism responsible for the observed change in magnetic order is usually a very complicated task[2-4]. In the case of absorption-induced thermal effects, the measured signals should be independent of the pulse polarization. Therefore, the dependence on the polarization of the excitation pulse is usually considered as a fingerprint of a non-thermal effect. However, since several different excitation mechanisms can coexist in many materials, their combination can lead to quite complex behavior. For example, the presence of magnetic circular dichroism (i.e. different absorption for different light helicities in a magnetic material) in all-optical helicity-dependent switching[3,10,11] leads to unequal heating of oppositely oriented magnetic domains for a given light helicity. Consequently, the resulting temperature gradient can lead to a helicity-dependent movement of the domain walls[11], i.e. to a polarization sensitivity of a fully thermal mechanism. Polarization sensitivity is therefore a necessary but not sufficient condition for a non-thermal excitation mechanism.

Although deeply fundamental in nature, ultrafast magnetism studies are also very important for various potential applications. For example, all-optical magnetization switching is considered the least-dissipative and fastest method for magnetic writing in memory applications[3,12]. In addition, the study of thin films of metallic magnets is attracting attention because these films are the basic building blocks of all spintronic devices, which are generally regarded as the most promising successors of current information technology devices[13-15]. The helicity-dependent effects are quite common in all conducting types of magnetic materials – metals[10,11,16-20], semiconductors[21-23], and insulators[24-26]. In contrast, effects that depend on the linear polarization of the laser pulses are usually only observed in insulating magnetic oxides[25-30]. So far, there is only one report describing such an effect in thin metal films – magnetization precession depending on the linear polarization of the pump pulses observed in a ferromagnetic Co-based heterostructure[31]. However, this is not a property of Co itself, but it is due to an optical rectification effect in the coupled ferroelectric $BiFeO_3$ layer[31]. Since metallic magnets strongly absorb light, the experimentally detected polarization-dependent magneto-optical signals used to reveal the magnetic dynamics are usually overlaid by a strong polarization-independent background[17,19,20,31,32], which makes their investigation considerably more difficult.

Currently, commercially available spintronic devices use thin films of ferromagnetic metals as the active material[33-35]. Antiferromagnets (AFs) are also very promising materials for future spintronic applications, as they combine many interesting properties[35-37]. In particular, the absence of net magnetization and stray fields eliminates crosstalk between adjacent devices, enabling high-density arrangement and making them robust to external magnetic fields. Furthermore, due to the strong exchange coupling between sublattices in AFs, the intrinsic resonance frequencies are in the terahertz frequency range, in contrast to the gigahertz frequencies in ferromagnets[37], which opens up the prospect of extremely fast device operation.

Experimentally, the absence of a net magnetic moment and effects linear in magnetization make the manipulation and probing of compensated collinear AFs very difficult[37,38]. For this reason, noncollinear AFs have started to attract considerable attention[39-43] as effects linear in magnetic moments – including the anomalous Hall effect[44], magneto-optical Kerr effect[45], and tunneling magnetoresistence[46], can exist in these materials (as in ferromagnets) despite a vanishingly small magnetization. Three types of noncollinear AFs with coplanar triangularly arranged Mn sublattice moments are typically studied in the literature – hexagonal Heuslers Mn$_3$X (X = Sn, Ge,...),[44,45,47,48] cubic antiperovskites Mn$_3$Y (Y = Ir, Pt,...)[46,49,50] and Mn$_3$AN (A = Ni, Ga, Sn,...)[51-53]. For spintronic applications, however, the exploration of the Mn$_3$X – type materials is by far the most common[39-43]. Accordingly, time-resolved pump-probe experiments have only been reported for Mn$_3$Sn, where only polarization-independent (i.e., thermal) demagnetization has been observed[54-56].

In this work, we report on pump-probe experiments in noncollinear antiferromagnetic Mn-based antiperovskite nitrides Mn$_3$NiN and Mn$_3$GaN. Using the magneto-optical (MO) Voigt effect, we observed an ultrafast (picosecond) pump-induced change in spin order. The measured MO signals are controlled by the orientation of the linear polarization of the pump pulses relative to that of the probe pulse. Moreover, the observed degree of polarization-dependence is extremely high, reaching ≈ 95% for probe pulses at normal incidence, a value unprecedented in metallic magnets. The performed symmetry analysis and microscopic calculations show that the laser-induced torque can in principle explain the observed MO signals, but for some polarizations there is a large discrepancy between theoretical predictions and experimental data. Therefore, it is very likely that the experimentally observed pump-induced change in magnetic order goes beyond the laser-induced torque, and we propose spin-spiral formation as a possible mechanism.

**RESULTS**

**Samples and experimental technique**

Films of antiferromagnetic metals Mn$_3$NiN and Mn$_3$GaN with a thickness of 13 and 40 nm, respectively, grown on (001)-oriented MgO substrates were investigated. In these materials, the magnetic frustration of the Mn atoms stabilizes the Kagome-like structure in the (111) plane. The resulting noncollinear antiferromagnetic spin structure can take two forms[57,58], known as $\Gamma^{4g}$ and $\Gamma^{5g}$, which are expected to predominate in Mn$_3$NiN and Mn$_3$GaN, respectively, as shown in Figs. 1a and 3a. (For more details on sample preparation and characterization see Methods and Supplementary Note 1.) The magnetic order changes induced by femtosecond laser pulses were investigated using the magneto-optical (MO) pump-probe technique[59] in a transmission geometry, as schematically shown in Fig. 1a. Here, the absorption of linearly polarized pump pulses with an orientation of the polarization plane $\alpha$ leads to a change in ellipticity and/or rotation, which is measured by time-delayed linearly polarized probe pulses with an orientation of polarization plane $\beta$. (For further details see Methods.)

**Magnetic dynamics controlled by linear polarization orientation in Mn$_3$NiN and Mn$_3$GaN**

In Figs. 1b and 1c we show the time-resolved change in ellipticity induced in Mn$_3$NiN by pump pulses with a polarization orientation $\alpha$, measured by probe pulses with a polarization orientation $\beta = 0°$ and 65°, respectively. The detected MO signal strongly depends on the linear polarization orientation of both the pump and probe pulses and almost disappears when the sample is excited by circularly polarized pump pulses $\sigma$. The detailed dependencies of the MO signal on the pump-polarization orientation are shown in Fig. 1d. For both probe-polarizations, the largest MO signal is measured when the angle between pump and probe polarization is 45° or 135°. Interestingly, the magnitude of the observed MO signal does not depend on the orientation of the pump-polarization with respect to the crystallographic directions in the sample (see also Supplementary Fig. S7b), which is quite surprising since the sample studied is a single-crystalline epitaxial film (as shown in Supplementary Fig. S1). As can be seen in Fig. 1e, the observed degree of linear polarization dependence (DLPD, see Methods) is ≈ 95%, which is very interesting as only polarization-independent (i.e. thermal) demagnetization has been experimentally observed in pump-probe experiments for metallic AFs so far[39]. The DLPD does not relax significantly within the duration of the pump-induced MO signal, which decays with a time constant of ≈ 2.3 ps (see Fig. 1b). The MO signals depend superlinearly on the pump intensity (see inset in Fig. 1b), but the DLPD does not seem to be significantly affected by the pump intensity (Supplementary Fig. S10).

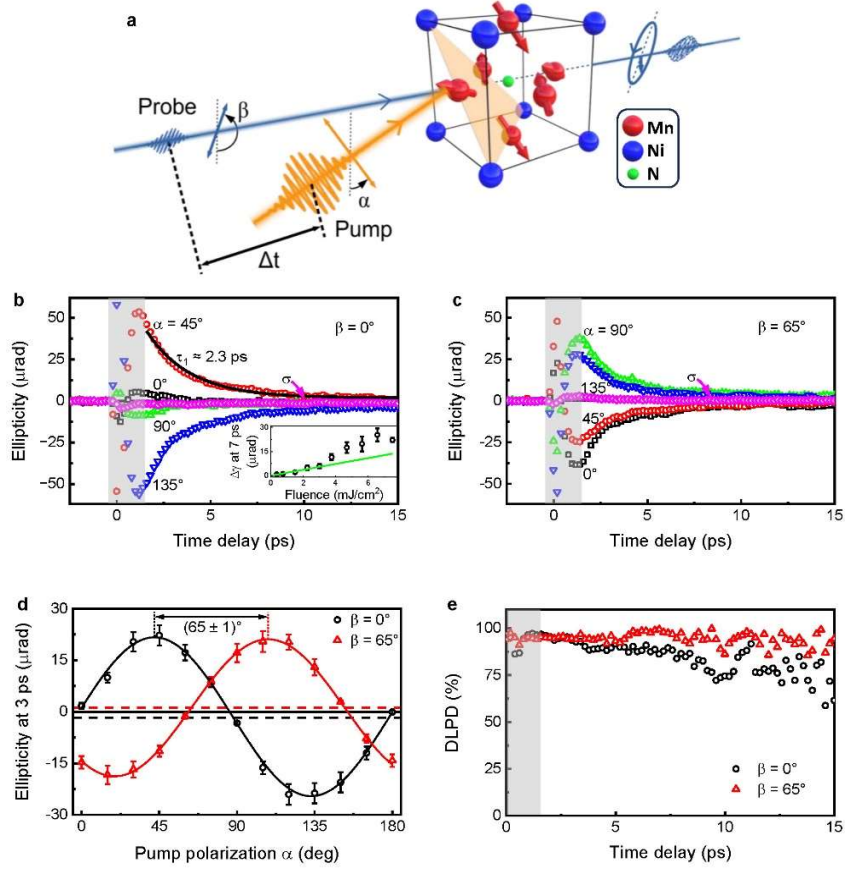

**Fig. 1: Magnetic dynamics controlled by the orientation of linear polarization of femtosecond laser pulses in Mn$_3$NiN.**
**a,** Schematics of the pump-probe experiment used to induce and detect magnetic dynamics in the noncollinear antiferromagnetic phase $\Gamma^{4g}$ of Mn$_3$NiN. The pump-induced change of magneto-optical (MO) signal (ellipticity and/or rotation) is measured simultaneously with the transmitted probe pulse intensity as a function of the time delay $\Delta t$ between the pump and probe pulses. The orientations of the linear polarization of the pump and probe pulses are described by angles $\alpha$ and $\beta$, respectively. **b,** Dynamics of ellipticity induced by pump pulses with various polarization orientations $\alpha$ measured by probe pulses with polarization orientation $\beta = 0°$ as a function of $\Delta t$ (points); the solid line is a monoexponential fit with time constant $\tau_1 \approx 2.3$ ps. Note that for circularly polarized pump pulses $\sigma$ the pump-induced MO signal is negligible. The gray area denotes the time window where the measured MO signals might not correspond to magnetic dynamics (cf. Fig. 2c). Inset: Measured intensity dependence of the ellipticity change (points), line depicts the linear dependence. **c,** Same as **b** for $\beta = 65°$. **d,** Pump-polarization dependence of ellipticity change at $\Delta t = 3$ ps measured for $\beta = 0°$ (black circles) and $\beta = 65°$ (red triangles). Solid lines are fits by harmonic functions with polarization-independent backgrounds (dashed horizontal lines), revealing the same amplitude and an angular shift of 65° in these dependencies. **e,** Degree of linear polarization dependence (DLPD) computed for $\beta = 0°$ (black circles) and $\beta = 65°$ (red triangles) from b and c. Sample temperature 25 K, pump fluence $\approx 3.8$ mJ/cm$^2$, pump wavelength 870 nm, and probe wavelength 435 nm. The error bars indicate the experimental uncertainty in MO signal determination from the measured pump-probe traces.

In principle, the measured dynamic MO signals can reflect not only the pump-induced change in magnetic order, but also the pump-induced change of a complex index of refraction (the so-called "optical part" of the MO signal)[60]. To verify the magnetic origin of the detected MO signals, we measured their temperature dependence, which is shown in Fig. 2a.

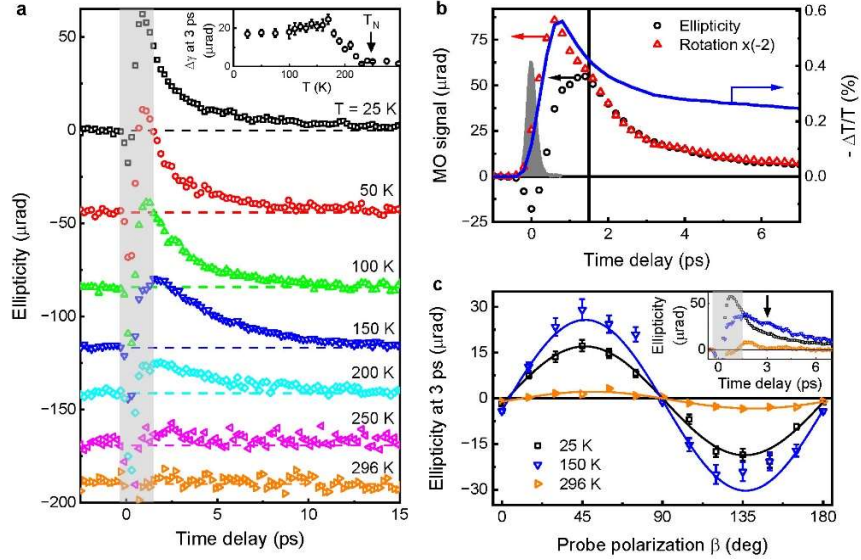

**Fig. 2: Magnetic origin of the magneto-optical signals observed in Mn$_3$NiN.**

**a,** Dynamics of ellipticity induced by pump pulses with polarization $\alpha = 90°$ measured by probe pulses with polarization $\beta = 45°$ at various base temperatures; data are vertically offset for clarity. Inset: Temperature dependence of the ellipticity change at $\Delta t = 3$ ps. The vertical arrow indicates the Néel temperature. **b,** Dynamics of ellipticity (black circles), rotation (red triangles) and transient transmission $\Delta T/T$ (blue line) at 25 K. The filled profile shows the pump-probe cross-correlation function, measured by sum-frequency generation in BBO crystal. The solid vertical line at $\Delta t = 1.5$ ps indicates the time delay up to which the measured MO signals do not correspond directly to magnetic dynamics; this time window is shown in gray in other figures. **c,** Probe-polarization dependence of ellipticity change at $\Delta t = 3$ ps at 25 K (black squares), 150 K (blue down triangles) and 296 K (orange right triangles). Solid lines are fits by harmonic functions with a temperature-independent phase. Inset: Dynamics of ellipticity for short time delays at the indicated temperatures, the vertical arrow marks $\Delta t = 3$ ps. Pump fluence ≈ 3.8 mJ/cm$^2$; pump (probe) wavelength 800 nm (400 nm) was used to measure the data shown in **a** and **c**, and 870 nm (435 nm) was used to measure data shown in **b**. The error bars indicate the experimental uncertainty in MO signal determination from the measured pump-probe traces.

The disappearance of the signals at the Néel temperature (see inset) is a clear indication of their connection with the magnetic order in the sample. In addition, the comparison of the dynamics of rotation and ellipticity shown in Fig. 2b shows that they coincide for time delays above ≈ 1.5 ps, further confirming the magnetic origin[60] of the observed MO dynamics in this time range (see also Supplementary Fig. S13 and the accompanying text). Further information can be obtained by comparing the dynamics of the MO signal with those of the transient transmission $\Delta T/T$, which on a sub-nanosecond time scale reflects the heat dissipation by which a laser-excited metal returns to equilibrium[61]. If the detected MO signals were related solely to the pump-induced thermal demagnetization of the material, the decay to equilibrium should be as slow as that of the $\Delta T/T$ signals[38]. The much longer and polarization-independent dynamics of $\Delta T/T$ (see Fig. 2b and Supplementary Fig. S8) show that this is not the case. Fig. 2c shows the dependence of the measured MO signal on the orientation of the probe-polarization, which is similar to that reported for other AFs[38,54,62]. The observed harmonic dependence is a clear indication that the Voigt effect (sometimes termed as the Cotton-Mouton effect), an MO effect that is quadratic in magnetic moment[63], dominates in the measured signals for the experimental geometry used, where the probe beam is perpendicular to the sample. When the probe beam is not perpendicular to the sample, MO effects that are linear in magnetic moment[64] also contribute to the measured signals[65]. This leads to a partial reduction of the DLPD, but apart from this, the pump-polarization sensitivity of the measured MO signals does not change (see Supplementary Figs. S9 and S10).

The experimentally observed decay of the pump-induced MO signal is extremely fast in the investigated as-grown $Mn_3NiN$ sample (see Fig. 1b). However, the derived characteristic time constant of $\approx 2.3$ ps does not appear to be an intrinsic property of $Mn_3NiN$. Recent study using high-resolution transmission electron microscopy[66] have shown that large local strain fields caused by lattice defects are present in $Mn_3NiN$ films grown on MgO (Supplementary Fig. S1). As we show in Supplementary Fig. S3, these strains can be strongly suppressed by post-growth annealing, which improves the crystallinity of the sample. While similar pump-polarization dependent MO signals are observed in the as-grown and annealed films (Supplementary Fig. S14), annealing leads to an increase in the time constants describing the decay of the MO signal and $\Delta T/T$ (Supplementary Fig. S14). Consequently, a more reasonable estimate of the intrinsic relaxation time of the magnetization order perturbed by the laser in $Mn_3NiN$ seems to be the time constant of $\approx 90$ ps derived for the annealed sample, which is still much shorter than typical magnetization relaxation times in ferromagnetically ordered materials[2-4].

To verify the "universality" of the observed phenomena, the same MO experiments were also conducted with another member of the antiperovskite nitride family, $Mn_3GaN$. This epilayer was prepared using a different growth technique and is 3-times thicker than the $Mn_3NiN$ layer (see Methods). Additionally, a different wavelength of the probe pulses and a different MO effect (rotation for $Mn_3GaN$ and ellipticity for $Mn_3NiN$, corresponding to the real and imaginary parts of the Voigt effect, respectively) were intentionally used in this control experiment. In the noncollinear antiferromagnetic phase $\Gamma^{5g}$ of $Mn_3GaN$ (Fig. 3a), all Mn magnetic moments are rotated by 90° in the (111) plane compared to the $\Gamma^{4g}$ phase of $Mn_3NiN$[57,58]. For this reason, the $\Gamma^{5g}$ phase does not exhibit the symmetry breaking required for the existence of effects such as the anomalous Hall[67] and Nernst[68,69] effects. Nevertheless, a very strong sensitivity to the linear polarization orientation of the pump pulses is also present in $Mn_3GaN$, see Fig. 3. Thanks to the higher Néel temperature in $Mn_3GaN$ ($T_N \approx 246$ K and 320 K in $Mn_3NiN$ and $Mn_3GaN$, respectively), this effect is observed in the sample even at room temperature (see inset in Fig. 3e and Supplementary Fig. S15).

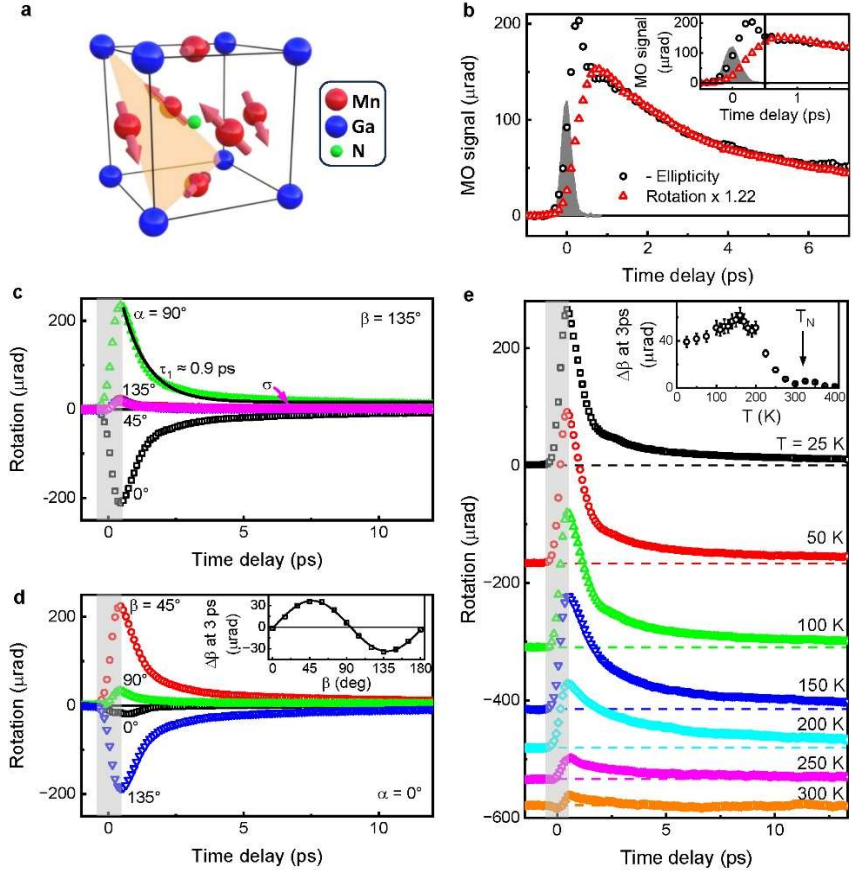

**Fig. 3: Magnetic dynamics controlled by the orientation of linear polarization of femtosecond laser pulses in $Mn_3GaN$.**
**a,** Noncollinear antiferromagnetic phase $\Gamma^{5g}$ of $Mn_3GaN$. **b,** Dynamics of pump-induced ellipticity (black circles) and rotation (red triangles); the filled profile shows the pump-probe cross-correlation function. Inset: MO signal dynamics around $\Delta t = 0$. The solid vertical line at $\Delta t = 500$ fs indicates the time delay up to which the measured MO signals do not correspond directly to magnetic dynamics; this time window is shown in gray in other figures. **c,** Dynamics of rotation induced by pump pulses with various polarization states measured by probe pulses with $\beta = 135°$ (points); the solid line is a monoexponential fit with time constant $\tau_1 \approx 0.9$ ps. **d,** Dynamics of rotation induced by pump pulses with $\alpha = 0°$ measured by probe pulses with various $\beta$. Inset: Probe-polarization dependence of rotation at $\Delta t = 3$ ps (points); the solid line is a harmonic fit. **e,** Dynamics of rotation at various base temperatures for $\alpha = 90°$ and $\beta = 135°$. Inset: Temperature dependence of rotation change at $\Delta t = 3$ ps. The vertical arrow indicates the Néel temperature. Pump fluence $\approx 6$ mJ/cm$^2$ and pump (probe) wavelength 820 nm (532 nm) were used; base temperature was 100 K in **b** and 25 K in **c** and **d**. The error bars indicate the experimental uncertainty in MO signal determination from the measured pump-probe traces.

**Laser-induced torque due to electrons photoexcited at neighboring sublattices**

While a DC electric field can produce a torque on the magnetic order[70], the electric field of the optical pulse oscillates much faster than the speed of the magnetic dynamics and is therefore not expected to cause a change in the magnetic order in the linear order. However, in the second order in the electric field, the torque may contain a time-independent component that can cause magnetic dynamics. This has been studied theoretically for circularly polarized light in ferromagnets[71-73] and linearly polarized light in a collinear antiferromagnet[74]. As we show by symmetry analysis (Supplementary Note 3) and microscopic calculations (Supplementary Note 4), the laser-induced torque can exist in antiperovskite nitrides. This is a consequence of noncollinear magnetic ordering, due to which spin is not conserved even in the absence of relativistic spin-orbit coupling. This is analogous to the nonrelativistic first-order current-induced torque recently studied in noncollinear antiferromagnets[75]. The nonrelativistic torque acting on a magnetic moment $\boldsymbol{M}_n$ in a sublattice $n$ for light polarization in the $(x,y)$-plane is given by an effective magnetic field $\boldsymbol{B}_n$ such that $\boldsymbol{T}_n = \boldsymbol{M}_n \times \boldsymbol{B}_n$ (see Supplementary Note 3 and Fig. 4a). Microscopically, the torque on one magnetic sublattice, e.g. 1, is given by the number of electrons excited from sublattices 2 and 3, whose spin orientation differs from that of sublattice 1 by the angle ±120°. Since the electron excitation rates for sublattices 2 and 3 are not the same (see Supplementary Note 4), the resulting spin-polarized population of excited electrons can be imagined as a spin current flowing to one sublattice from the other sublattices. The torques depicted in Fig. 4a, which are analogous to the antidamping-like spin-transfer torque in ferromagnets[76], are odd in the time-reversal ($T$-odd in the following). In principle, also torques even in the time-reversal ($T$-even) can exist in Mn$_3$NiN, see Supplementary Tables S3.1 and S3.2 for their form in all 8 possible magnetic domains[77] of the $\Gamma^{4g}$ phase of Mn$_3$NiN shown in Supplementary Fig. S18. However, simulations of the magnetic dynamics by the Landau-Lifschitz-Gilbert (LLG) equation show that $T$-even torques have only a very small effect on the magnetic state (Supplementary Fig. S23) because the effective fields drive the magnetic moments against the exchange interaction, which is very strong. In contrast, $T$-odd torques, which act against the anisotropy, can significantly alter the equilibrium magnetic state and lead to a considerable tilt out of the moments plane, as shown in Fig. 4b. For the modified magnetic state resulting from the LLG simulations, the permittivity tensor was calculated using linear response theory and the MO signal detected by the probe pulses was derived using the Yeh's formalism for all magnetic domains (Supplementary Note 4). The obtained probe-polarization dependence for the pump polarization $\alpha$ = 0°, shown in Fig. 4a, is in perfect agreement with the measured data (Supplementary Fig. S7b).

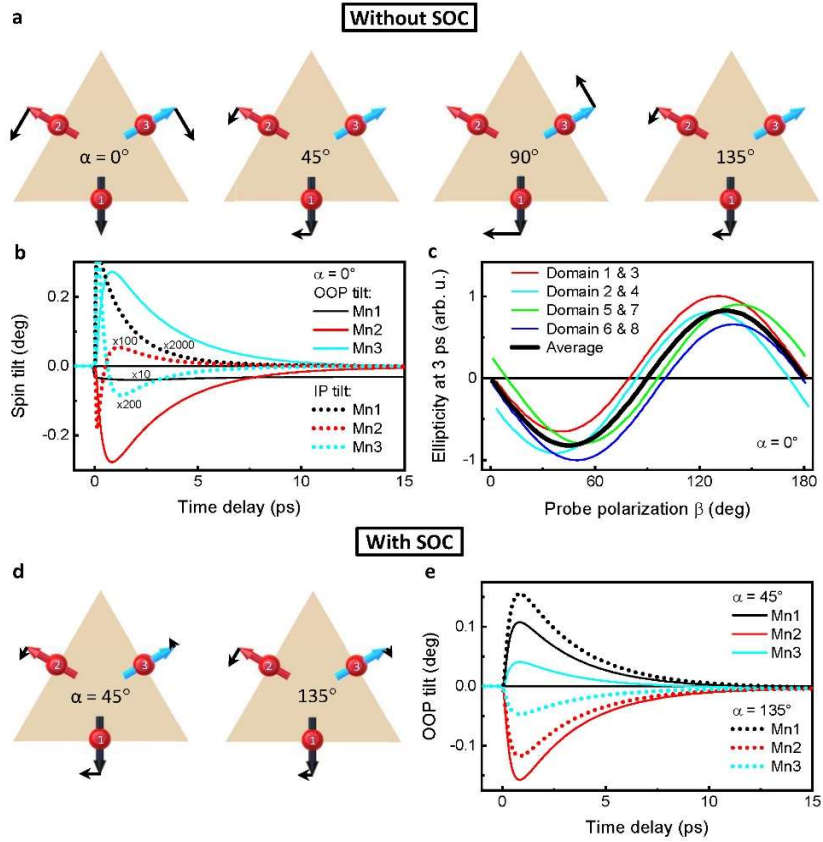

**Fig. 4: Laser-induced torque due to electrons photoexcited at neighboring sublattices.**
**a,** Torques (black arrows) induced on the three spin sublattices by pump pulses with the indicated linear polarizations $\alpha$ when spin-orbit coupling (SOC) is neglected. Red circles represent Mn atoms and the corresponding arrows indicate their spins. The torque directions are determined by both symmetry analysis (Supplementary Note 3) and microscopic calculations (Supplementary Note 4). The shown torques correspond to domain 1 in Supplementary Table S3.1 and to Supplementary Eqs. S4.4 and S4.7. **b,** Dynamics of the out-of-plane (OOP) and in-plane (IP) Mn spin tilts calculated by the LLG equation for $\alpha$ = 0° without SOC. Note that the OOP tilt is much larger than the IP tilt. Calculations are for magnetic domain 1 (see Fig. S18 in Supplementary Note 4 for domain visualization). **c,** Probe-polarization dependence of ellipticity change computed for the MO Voigt effect experienced by probe pulses at normal incidence from the spin tilts shown in **b** at $\Delta t$ = 3 ps for all 8 magnetic domains of the $\Gamma^{4g}$ phase. **d,** As in **a** with SOC included. **e,** Calculated OOP Mn spin tilt with SOC for $\alpha$ = 45° and 135° in domain 1.

When the orientation of the pump-polarization is rotated, the overall spin polarization of the excited electrons changes (Supplementary Fig. S21), leading to a torque change, as shown in Supplementary Table S3.1 and Fig. 4a. However, there is a large discrepancy between the theoretical predictions and the experimental results for $\alpha$ = 45° and 135°. While symmetry requires the same torques and thus the same MO signals for these two polarizations (see Fig. 4a), opposite MO signals are observed experimentally (Fig. 1b). When spin-orbit coupling is taken into account, the symmetry is significantly reduced and the torques obtained for $\alpha$ = 45° and 135° are no longer the same, see Fig. 4d and Supplementary Notes 3 and 4. However, the calculations show only a slight change in the moment dynamics when spin-orbit coupling is included (Fig. 4e), which cannot explain the experimental observations. In principle, this discrepancy with the experiment could be caused by a symmetry-breaking non-ideality of the investigated samples. Nevertheless, the detailed structural characterization by X-ray diffraction excludes a polycrystallinity of the sample (Supplementary Fig. S1). Similarly, the randomly oriented local strain fields that exist in the $Mn_3NiN$ samples[66] do not seem to play a major role, as these strains are strongly suppressed in the annealed $Mn_3NiN$ sample (cf. Supplementary Figs. S1 and S3), but opposite MO signals are also observed here for $\alpha = 45°$ and $135°$ (Supplementary Fig. S14). Therefore, it is very likely that the experimentally observed pump-induced change in magnetic order goes beyond the laser-induced torque, as described below.

**Laser-induced formation of spin spiral**

The formation of the skyrmion phase by ultrafast laser pulses has been observed experimentally in thin-film ferromagnets exhibiting chiral interactions[7]. Recently, it has also been proposed theoretically that ultrafast laser pulses can generate chiral magnetic textures, such as spin spirals or meron-antimeron pairs, even in systems that do not exhibit chiral magnetic interactions at equilibrium[78-80]. It is therefore possible that the MO signals observed in the experiment are due to the formation of magnetic textures such as spin spirals. Simulating such an effect requires solving the time-dependent Schrödinger equation for combined electronic and magnetic dynamics[78-80]. This creates a spin-orbit-like interaction that forms a Dzyaloshinskii-Moriya-type interaction resulting in the establishment of a spin spiral in the system[78,80]. Performing these calculations for a realistic model of $Mn_3NiN$ and $Mn_3GaN$, including the influence of existing magnetic domains, is beyond the scope of this work. However, our preliminary results for a noncollinear 2D model based on this technique[78,80], described in Supplementary Note 5, show that linearly polarized femtosecond laser pulses can generate spin spirals along the polarization directions (see Fig. 5 and Supplementary Video 1). The induced metastable helical pattern is formed by three sublattice spirals with the same wavelength but a constant phase difference, see Fig. 5c. The generated spin spirals are isotropic with respect to the crystal orientation (Supplementary Fig. S28), which is consistent with the experimental observations. Moreover, these spiral patterns can be generated for both the $\Gamma^{4g}$ and $\Gamma^{5g}$ noncollinear antiferromagnetic spin structures, as shown in Supplementary Fig. S28 and Supplementary Video 2. Finally, the computed intensity dependence (Supplementary Fig. S27g) is similar to the intensity dependence measured experimentally (inset in Fig. 1b).

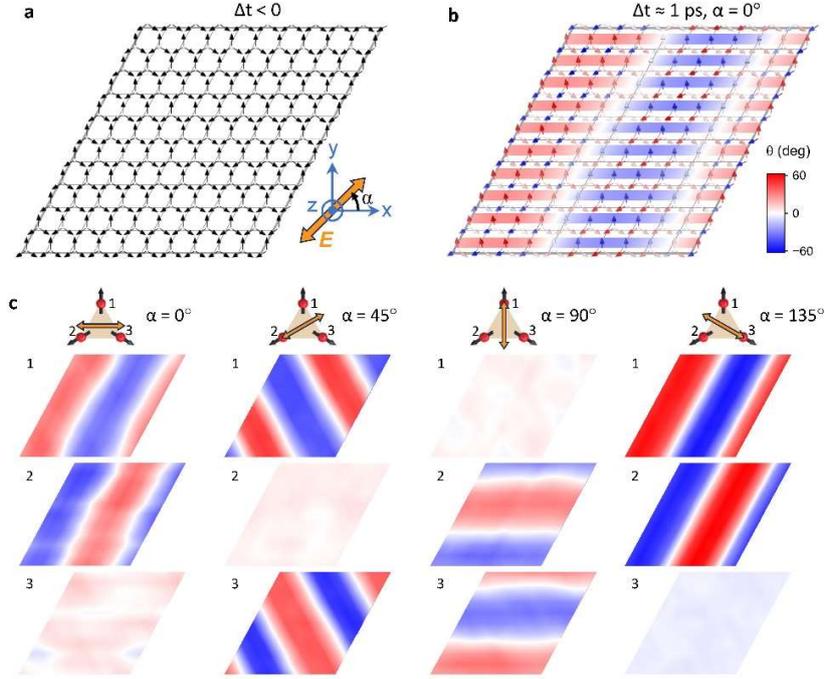

**Fig. 5: Laser-induced formation of a spin spiral for a noncollinear 2D model.**
**a,** 10 × 10 array of the (111)-plane of the non-collinear $\Gamma^{4g}$ unit cell before excitation by a pump pulse, along with the corresponding coordinate system. Pump pulses are incident along the $z$ axis and are polarized in the $(x,y)$-plane along direction $\alpha$ (double-headed orange arrow). **b,** Spin structure ≈ 1 ps after excitation by pump pulse with polarization $\alpha$ = 0°. The coloring of individual spins (arrows) corresponds to the pump-induced $z$-component of the magnetic moments. **c,** Contour graphs of induced $z$-component of magnetic moments for each of the three sublattices for the depicted pump polarizations $\alpha$.

## DISCUSSION

To sum up, the experimentally observed pump-induced magneto-optical (MO) signals in metallic Mn-based antiperovskite nitrides indicate a change in spin order controlled by the orientation of linearly polarized pump pulses. The magnitude of the pump-polarization-dependent MO signals detected with a probe polarization $\beta$ is largest when the orientation of the pump polarization $\alpha$ is equal to $\beta \pm 45°$. Additionally, the MO signal changes sign when either the pump or probe polarization is rotated by 90°, but the MO signal does not depend on the crystallographic directions in the studied epitaxial samples (Fig. 1d and Supplementary Fig. S7b). Such MO signals are observed over a wide range of excitation intensities and wavelengths of the pump and probe pulses in samples of Mn$_3$NiN and Mn$_3$GaN, in which different noncollinear antiferromagnetic spin structures $\Gamma^{4g}$ and $\Gamma^{5g}$ should prevail.

The performed symmetry analysis (Supplementary Note 3) and microscopic calculations (Supplementary Note 4) show that the laser-induced torque can in principle explain the observed MO signals, but there is a large discrepancy between the theoretical predictions and the experimental data for $\alpha = \beta + 45°$ and $\beta - 45°$. While symmetry requires the same torques and thus the same MO signals for these two polarizations (Fig. 4a and Supplementary Table S3.1), opposite MO signals are observed experimentally (Fig. 1b). When spin-orbit coupling is included in the calculations, the symmetry is significantly reduced and the torques obtained for $\alpha = \beta + 45°$ and $\beta - 45°$ are no longer equal (Fig. 4d), but the resulting small change in the moments dynamics cannot explain the experimental observations. Although our calculations are based on a simplified model that does not include all possible mechanisms, the symmetry analysis (Supplementary Note 3) is general and describes any second-order optically induced torque, including the inverse Faraday and Cotton-Mouton MO effects[8,9]. The discrepancy between theory and experiment could be due to mechanisms that were not considered in our model. However, such a mechanism must be relativistic, since the non-relativistic symmetry is not consistent with the experimental data. While we cannot exclude this possibility, the experimental observation of exactly opposite MO signals for the polarization directions $\beta + 45°$ and $\beta - 45°$ suggests a different explanation, since relativistic symmetry does not impose any constraints on these directions.

The preliminary calculations of laser pulse-induced generation of chiral magnetic textures[78,80] beyond the laser-induced torque appear to be consistent with the experimental observations, including the isotropy with respect to crystal orientation (Fig. 5c and Supplementary Fig. S28) and the existence of the effect for both the $\Gamma^{4g}$ and $\Gamma^{5g}$ noncollinear antiferromagnetic spin structures (Supplementary Fig. S28 and Supplementary Video 2). Note that this mechanism differs from the reported skyrmion nucleation by femtosecond lasers[7] where the laser is used only as a thermal stimulus to drive the system to its ground state[5-7]. In our experiment, the laser generates coherent interactions between different quantum states that bring the system to a metastable state. The fact that the spin spiral follows the polarization of the laser indicates that the interaction responsible for the spiral formation originates from the laser and not from the material itself.

In conclusion, our work opens a new avenue for ultrafast manipulation of magnetization dynamics in metallic noncollinear antiferromagnets by linearly polarized light. Nevertheless, further theoretical and experimental work is needed before the exact origin and full potential of this non-thermal control of magnetic order can be revealed.

**ONLINE CONTENT**

Any methods, additional references, Nature Portfolio reporting summaries, source data, extended data, supplementary information, acknowledgements, peer review information; details of author contributions and competing interests; and statements of data and code availability are available at [to be completed].

**ONLINE METHODS**

**Samples**

Epilayers of Mn$_3$NiN with a thickness of 13 nm were grown by the pulsed laser deposition (PLD) technique using a Mn$_3$NiN target at 500°C in $1.2 \times 10^{-4}$ Torr N$_2$ partial pressure[66]. The annealed Mn$_3$NiN film was prepared by heating the as-grown film in $1.2 \times 10^{-4}$ Torr N$_2$ partial pressure to 700°C at a heating rate 20 K/min, and then cooling it to room temperature at 10 K/min. An epilayer of Mn$_3$GaN with a thickness of 40 nm was grown by the physical vapour deposition (PVD) method from a MnGa target at 350°C under $2.5 \times 10^{-8}$ Torr at 8% N$_2$ partial pressure in an Ar atmosphere.

X-ray diffraction (XRD) measurements using a Malvern PANalytical Empyrean diffractometer indicated a good crystal quality of the as-grown Mn$_3$NiN sample and measurements using a Smartlab rotating anode confirmed the single-crystallinity of the film. The unit cell exhibited a net tetragonality $c/a = 0.992$. Strain field mapping, using a combination of scanning transmission electron microscopy (STEM) and geometric phase analysis (GPA), revealed a large dislocation network, which creates strong local strain fields. For a magnetic characterization, we used a Quantum Design magnetic property measurement system (SQUID) magnetometer. From zero-field cooled (ZFC) and field-cooled warming (FCW) measurements we extracted the Néel temperature $T_N = 246$ K of the as-grown film. Hysteretic anomalous Hall effect (AHE) curves obtained by sweeping the applied magnetic field at various temperatures confirmed the appearance of the $\Gamma^{4g}$ non-collinear magnetic phase below $T_N$.

XRD characterization of the annealed Mn$_3$NiN film using a Malvern PANalytical Empyrean diffractometer revealed a higher degree of crystallinity compared to the as-grown film. This is also supported by strain field mapping, which shows a lower density of dislocations and weaker local strain fields. The unit cell exhibited a net tetragonality $c/a = 0.997$. ZFC and FCW magnetometry measurements revealed $T_N = 242$ K.

For structural characterization of the Mn$_3$GaN film, we performed XRD measurements, which indicated a crystal quality similar to that of the as-grown Mn$_3$NiN film. Magnetic characterization using a vibrating-sample magnetometer (VSM) showed that the sample was magnetically ordered even at room temperature.

Further details about the sample characterization are provided in Supplementary Note 1.

**Pump-probe experiment**

A pump-probe experiment was performed using a Ti:sapphire oscillator (Mai Tai, Spectra Physics) as a source of ≈ 150 fs laser pulses. For the experiments in $Mn_3NiN$, the second harmonic of the oscillator output was generated in a BBO nonlinear crystal. The laser beams at fundamental and doubled photon energy were separated using a dichroic mirror and used as pump and probe beams, respectively. The experiments at 25 K were performed using the 80 MHz repetition rate of the oscillator and the temperature dependence (Figs. 2a and 2c) was measured with the repetition rate reduced to 8 MHz using a pulse picker. For the experiments in $Mn_3GaN$, the output of the oscillator at 820 nm with a repetition rate of 80 MHz was used as the pump beam and the output of an optical parametric oscillator (Inspire, Spectra Physics) at 532 nm was used as the probe beam. Both beams were focused onto the sample by lenses, resulting in a beam diameter of ≈ 15 μm and ≈ 4 μm (full width at half maximum, FWHM) for the pump and probe beams, respectively. The experiment was performed with pump pulses with a fluence from ≈ 0.5 to 7.5 mJ/cm², the fluence of the probe pulses was ≈ 0.2 mJ/cm². The sample was mounted on a cold finger of the closed-loop helium cryostat (ARS), where the temperature can be changed from 15 to 800 K. All experiments were performed on samples field-cooled[47,77] from room temperature to 25 K with an external magnetic field (generated by an electromagnet) of 530 mT applied in the sample plane.

The pump pulses were linearly polarized, with the orientation of the polarization plane controlled by a half-wave plate and described by an angle $\alpha$ (see Fig. 1a). For reference, the circularly polarized pump pulses $\sigma$ generated by a quarter-wave plate were used. The probe pulses were linearly polarized, with the orientation of the polarization plane controlled by a half-wave plate and described by an angle $\beta$. The dynamics of the MO signal, i.e. the pump-induced change in polarization rotation or ellipticity measured by probe pulses transmitted through the sample, was obtained by taking the *difference* of signals measured by the detectors in an optical bridge detection system[59]. Simultaneously with the MO signal, the *sum* of the signals from the detectors was also detected, corresponding to a change in probe intensity due to the pump-induced modification of the sample transmission[59]. The differential transmission $\Delta T/T = (T_E - T) / T$, where $T_E$ ($T$) is the transmission with (without) the pump pulse, was used as a measure for the transmission changes. The angle between the pump and probe beam was set to $\approx 15°$ and the external magnetic field was applied in a direction perpendicular to the probe beam. In our experimental setup, the sample orientation with respect to the laser beams could be varied, resulting in angles of incidence of the probe and pump beams labeled as AOI$_{probe}$ and AOI$_{pump}$, respectively. The experimental configuration corresponding to AOI$_{probe} \approx 0°$ (when AOI$_{pump} \approx 15°$) was used to measure all experimental data shown in Figures 1-3. The results obtained for other values of AOI$_{probe}$ are shown in Fig. S9 in Supplementary Note 2 and will be addressed in detail elsewhere[65].

To characterize how much the ellipticity change $\Delta\gamma$ (or, similarly, the rotation change $\Delta\beta$), measured by probe pulses with polarization $\beta$, depends on the linear polarization of pump pulses $\alpha$, we introduce a quantity termed "degree of linear polarization dependence" (DLPD)

$$\text{DLPD}(\alpha) \equiv 100\% \cdot |\Delta\gamma_\alpha| / (|\Delta\gamma_\alpha| + |\Delta\gamma_\sigma|). \tag{1}$$

Here, $\Delta\gamma_\alpha$ represents the measured signal for pump linear polarization $\alpha$ and $\Delta\gamma_\sigma$ is the signal measured for circularly polarized pump pulses $\sigma$, which corresponds to the pump-polarization-independent part of the MO signal (see also Fig. S5 and the coresponding text in Supplementary Note 2). For all measured data, DLPD($\alpha$) is largest when $\alpha - \beta \approx 45°$ or $135°$ and smallest when $\alpha - \beta \approx 0°$ or $90°$ (see Fig. S6 in Supplementary Note 2). The curves displayed in Fig. 1d are the averages of DLPD(45°) and DLPD(135°) for $\beta = 0°$, and DLPD(105°) and DLPD(15°) for $\beta = 65°$, respectively.

**DATA AVAILABILITY**

Data reported in this paper are available in the Zenodo repository [PERSISTENT WEB LINK TO DATASETS to be provided in Proofs] and are publicly available as of the date of publication.

**ACKNOWLEDGEMENTS**

We would like to thank Dominik Kriegner and Satya Prakash for X-ray diffraction characterization of Mn$_3$NiN sample and Martin Veis for magnetic characterization of Mn$_3$GaN sample. This work was supported by TERAFIT project No. CZ.02.01.01/00/22_008/0004594 funded by Ministry of Education Youth and Sports of the Czech Republic (MEYS CR), programme Johannes Amos Comenius (OP JAK), call Excellent Research, and by CzechNanoLab Research Infrastructure supported by MEYS CR (LM2023051), and by e-INFRA CZ (ID:90254) project of MEYS CR. K.C. acknowledges the support by Czech Science Foundation, Grant No. 23-04746S, J.Ž. and J.Z. acknowledge the support by Czech Science Foundation, Grant No. 25-18244S, J.Ž. thanks Dioscuri Program LV23025 funded by MPG and MEYS CR, S.G. acknowledges the support by the European Union (Physics for Future - Grant Agreement No. 101081515), and F.J. is grateful for support from the Royal Commission of 1851 Research Fellowship.

**AUTHOR CONTRIBUTIONS**

P.N. and J.K. planned the study. J.K. and M.N. carried out the pump-probe measurements and performed the data analysis with the help of P.N., E.S. and L.N.. F.J., F.R-B. and D.B. fabricated and characterized the Mn$_3$NiN samples. A.P.M, B.Z., X.S. and F.Y. fabricated and characterized the Mn$_3$GaN sample. K.C., S.G., J.Ž., T.O. and J.Z. developed the theoretical models and performed calculations. P.N. and J.K. wrote the manuscript with contributions from all authors.

**COMPETING INTERESTS**

The authors declare no competing interests.

**ADDITIONAL INFORMATION**

**Supplementary information**

The online version contains supplementary material available at [to be provided]. Supplemental materials consist of Supplementary Note 1. Sample characterization (containing Figures S1-S4), Supplementary Note 2. Additional time-resolved data (containing Figures S5-S16), Supplementary Note 3. Symmetry analysis (containing Tables S3.1-S3.4), Supplementary Note 4. Laser-induced torque due to electrons photoexcited from neighbouring sublattices (containing Figures S17-S25 and Table S4.1) and Supplementary Note 5. Real time quantum-classical dynamics of non-collinear magnet and formation of spin spiral (containing Figures S26-S31), Supplementary references, Supplementary Video 1. Generation of spin spiral for $\Gamma^{4g}$ spin structure by polarizations 135° and 45°, and Supplementary Video 2. Generation of spin spiral for $\Gamma^{5g}$ and $\Gamma^{4g}$ spin structures.

**Correspondence and requests for materials** should be addressed to Petr Němec.

# Ultrafast control of spin order by linearly polarized light in noncollinear antiferromagnetic metals: Supplementary information

J. Kimák[1], M. Nerodilová[1], K. Carva[1], S. Ghosh[2], J. Železný[2], T. Ostatnický[1],
J. Zemen[3], F. Johnson[4], D. Boldrin[5], F. Rendell-Bhatti[5], B. Zou[6], A.P. Mihai[6],
X. Sun[7], F. Yu[7], E. Schmoranzerová[1], L. Nádvorník[1], L. F. Cohen[8], P. Němec[1]

[1]Faculty of Mathematics and Physics, Charles University, Ke Karlovu 3, 121 16, Prague 2, Czech Republic.
[2]Institute of Physics ASCR, v.v.i., Cukrovarnická 10, 162 53, Prague 6, Czech Republic.
[3]Faculty of Electrical Engineering, Czech Technical University in Prague, Technická 2, 166 27, Prague 6, Czech Republic.
[4]Cavendish Laboratory, University of Cambridge, JJ Thomson Ave, CB3 0HE, Cambridge, UK.
[5]SUPA, School of Physics and Astronomy, University of Glasgow, G12 8QQ, Glasgow, Scotland, UK.
[6]LoMaRe Technologies Ltd, 6 London Street, SW7 2AZ, London, UK.
[7]LoMaRe Chip Technology Changzhou Co., Ltd., 400 Hanjiang Road, Changzhou, China.
[8]Department of Physics, Blackett Laboratory, Imperial College London, SW7 2AZ, London, UK.

# Contents

# 1 Supplementary Note 1. Sample characterization

Time-resolved pump-probe experiments were performed in 13 nm thick films of $Mn_3NiN$ and 40 nm thick film of $Mn_3GaN$, all grown on 0.5 mm thick (001)-oriented MgO substrates. In this part of the supplementary material we provide a detailed characterization of these films.

## 1.1 As-grown $Mn_3NiN$ film

### Structural and optical characterization

The as-grown film of $Mn_3NiN$ was grown at 500°C under 12 mTorr $N_2$ partial pressure by the Pulsed Laser Deposition technique. The bulk lattice parameter of the $Mn_3NiN$ target used for growth was 3.8805 Å.

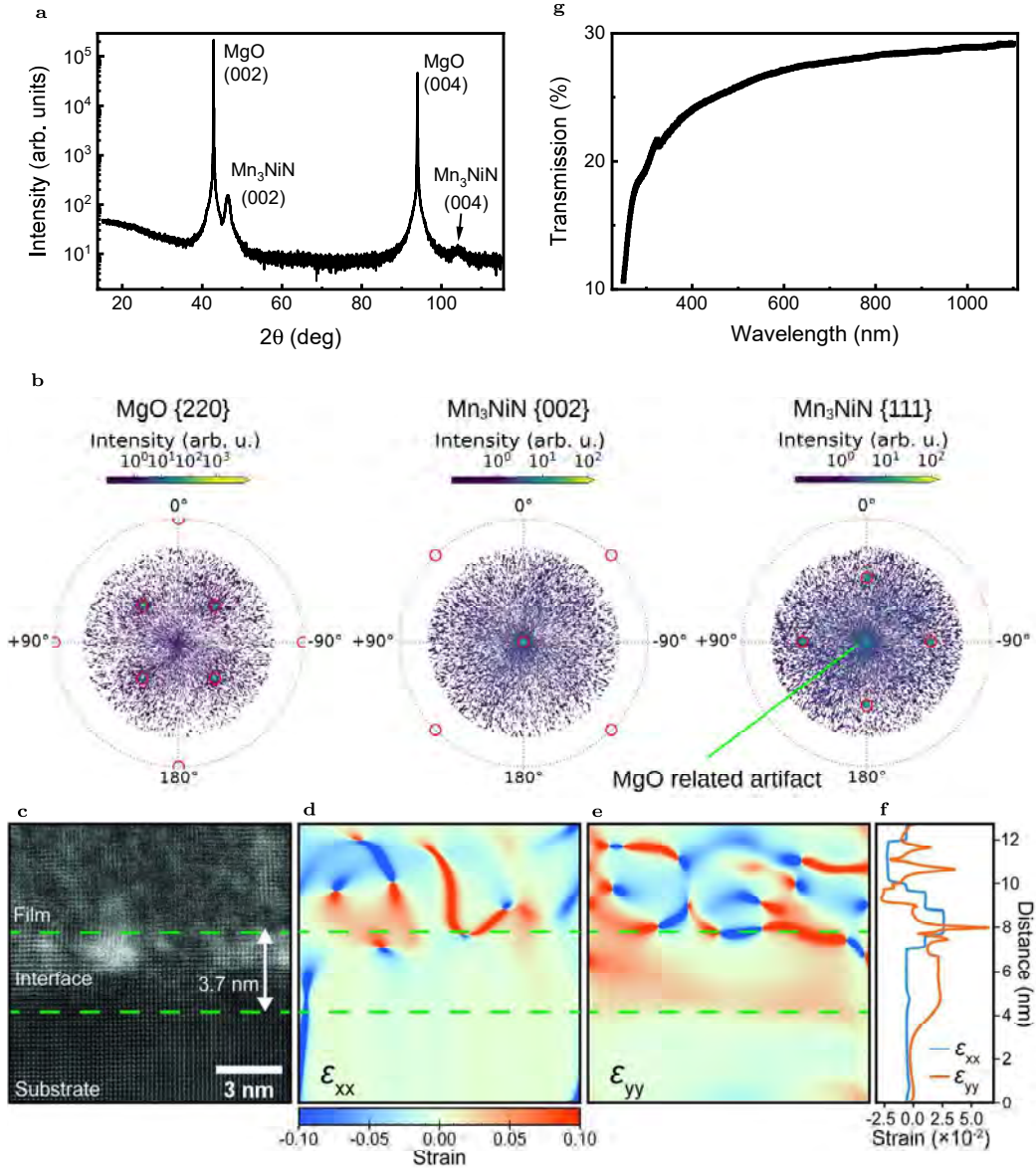

**Fig. S1**: **$Mn_3NiN$ as-grown sample structural and optical characterization. a**, XRD scan. **b**, Pole figures showing orientation distribution of lattice planes. Red circles mark positions for single crystalline orientation. **c**, Annular dark-field image from Scanning transmission electron microscope (STEM) of the $Mn_3NiN$ film, interfacial region and the MgO substrate. **d,e**, Strain maps of the $\varepsilon_{xx}$ and $\varepsilon_{yy}$ components of the strain tensor revealing a network of dislocations within the film. **f**, $\varepsilon_{xx}$ and $\varepsilon_{yy}$ averaged over rows of atoms. **g**, Spectrum of optical transmission of the sample measured at room temperature.

For structural characterization, X-ray diffraction (XRD) was performed using a Malvern Panalytical Empyrean diffractometer and a Smartlab rotating anode. The out-of-plane (OOP) XRD (see Fig. S1a) shows no other spurious or crystalline phases. From the scan we also extract the lattice parameter $c = 3.908$ Å, and from the Reciprocal space mapping (RSM) around the 113 peak (not shown) the lattice parameter $a = 3.940$ Å. The sample unit cell therefore shows a net tetragonality $c/a = 0.992$, which should lead to the development of a small net moment due to the piezomagnetic effect and an increase of $T_N$ [1–5]. The XRD-pole figure measurements (see Fig. S1b) at the Bragg angles corresponding to the MgO (220), Mn$_3$NiN (002) and Mn$_3$NiN (111) reflections further confirm the single-crystallinity of the film as other orientations of crystalline phases would be clearly visible here.

As shown in detail in [4], edge dislocation networks resulting in local strain fields may be present in thin films of Mn$_3$NiN. To address this, we performed Strain Field Mapping using the methods described in [4]. Fig. S1c shows images of Mn$_3$NiN film, a 3.7 nm thin mixed-phase interface and MgO substrate measured using scanning transmission electron microscopy (STEM). The film had a layer of Pt deposited on top (not shown), which helps the ion milling process. To extract the local strain fields, we used Bragg filtering generated by masking 100 and 001 spots in the Fourier transform combined with Geometric phase analysis (GPA), which was performed using the Strain++ software [6], which implements the algorithms outlined in [7]. Resulting strain fields in the in-plane (IP) direction $\varepsilon_{xx}$ and in the OOP direction $\varepsilon_{yy}$ are shown in Fig. S1d and S1e, respectively. The amplitude of the strain fields within the scanned area reaches values as large as $\approx 8 \times 10^{-2}$. Similarly to thin film of Mn$_3$Cu$_{0.5}$Sn$_{0.5}$N grown on MgO (Fig. 1 in [4]), the lattice of Mn$_3$NiN is closely matched to the substrate lattice IP in the interfacial area, but the OOP strain shows an expansion. This is even more clearly visible in the average over rows of atoms in Fig. S1f.

To characterize the optical response of the sample, we measured the optical transmission spectrum. Fig. S1g shows that the transmission of the film has a relatively simple dependence on the wavelength with no pronounced absorption or interference features.

## Magnetic characterization

Magnetic characterization measurements were performed using a Quantum Design Magnetic Property Measurement System SQUID magnetometer. To extract the Hall component, four terminal magnetotransport data were collected using the van der Pauw method and antisymmetrised. The film was mounted on quartz holders using low-temperature adhesive, such that the applied field was parallel to the field surface.

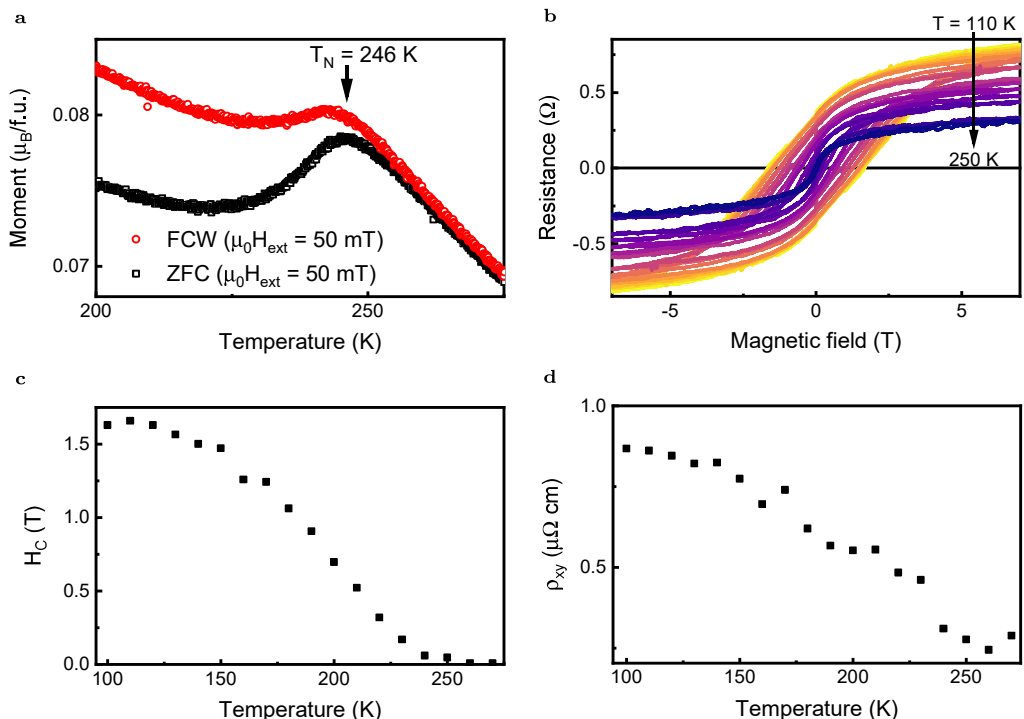

**Fig. S2**: **Mn$_3$NiN as-grown sample magnetic characterization.** **a** Field-cooled warming (FCW) and Zero-field cooled (ZFC) measurements of magnetic moment in applied magnetic field $\mu_0 H_{ext} = 50$ mT in the IP direction. The vertical arrow indicates the Néel temperature. **b**, Resistance as a function of applied IP magnetic field measured at various temperatures. **c**, Temperature dependence of coercive field. **d**, Temperature dependence of anomalous Hall resistivity.

From Zero-field cooled (ZFC) and Field-cooled warming (FCW) magnetometry measurements (see Fig. S2a), we extract the Néel temperature $T_N = 246$ K, which is, as expected from the XRD measurement, a bit larger than in films with $c/a > 1$ [1]. Hysteresis loops from above $T_N$ down to $T = 110\,\text{K}$ (see Fig. S2b) show a clear opening once the material enters the antiferromagnetic phase. To show this more clearly, we plot the temperature dependence of the coercive field $H_C$ and the anomalous Hall resistivity $\rho_{xy}$ in Fig. S2c and S2d, respectively. This is in agreement that the Mn$_3$NiN is in the non-collinear $\Gamma^{4g}$ phase below the $T_N$.

In general, Mn$_3$NiN could adopt two non-collinear triangular antiferromagnetic structures $\Gamma^{4g}$ and $\Gamma^{5g}$ [8]. Although both structures are fully compensated, it has been predicted that $\Gamma^{4g}$ possesses the required symmetry-breaking necessary to show Berry-curvature enhanced anomalous Hall effect (AHE) [9] and this has subsequently been measured in Mn$_3$NiN thin films [5, 10, 11]. Neutron diffraction on thin films of Mn$_3$NiN has also shown a $\Gamma^{4g}$ component below $T_N$ [5]. Conversely, $\Gamma^{5g}$ does not show the required symmetry breaking, therefore does not permit the AHE and the measured amplitude of magnetic moment could not solely explain the amplitude of AHE [9, 11].

## 1.2 Annealed Mn$_3$NiN film

The method of growth of the second Mn$_3$NiN film was exactly the same as of the first one, described in detail in the previous section. However, the key difference between these two films is the postgrowth annealing. The annealed Mn$_3$NiN film was heated in 12 mTorr N$_2$ partial pressure to 700°C at 20 K/min heating rate, and then cooled to room temperature at 10 K/min.

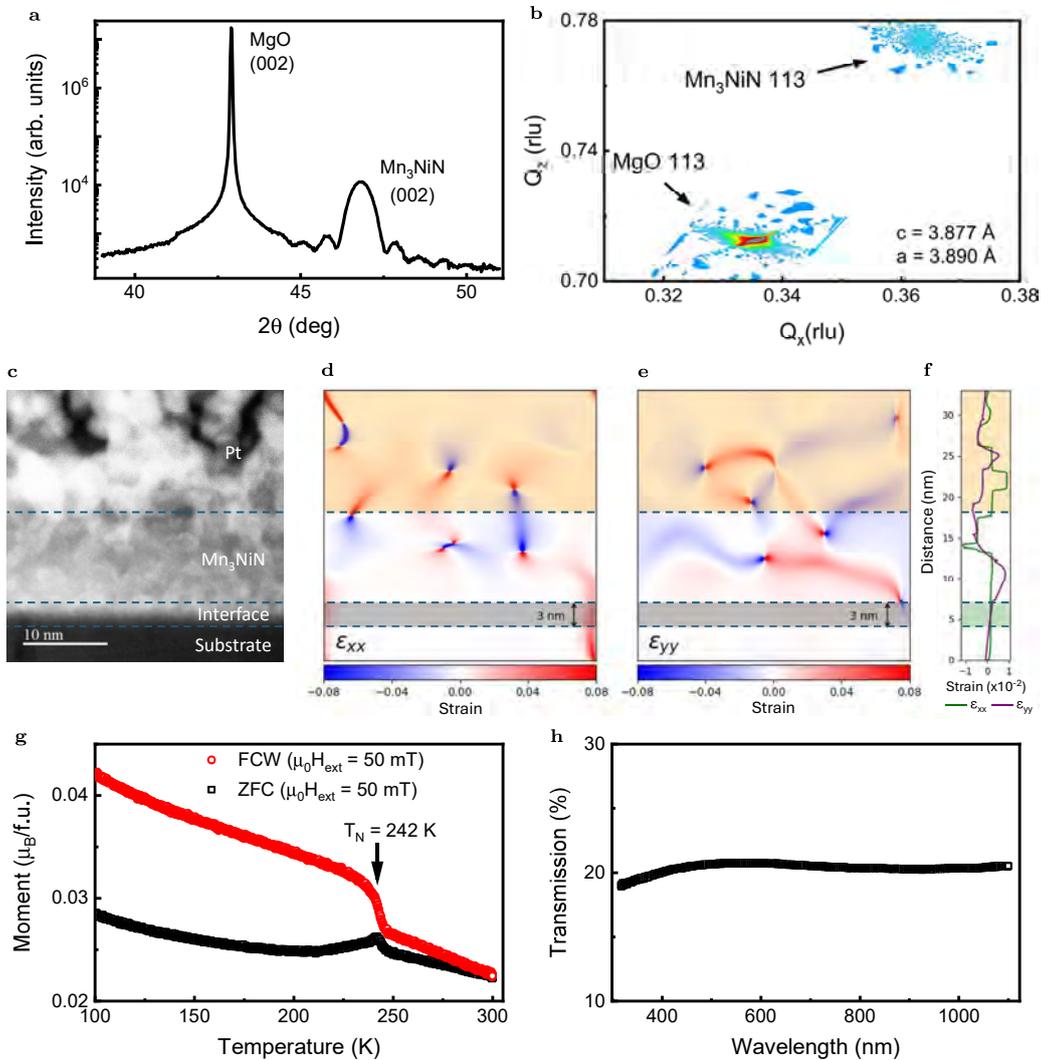

**Fig. S3**: **Mn$_3$NiN annealed sample characterization. a** XRD scan. **b**, RSM around the 113 reflection, from which we extract lattice constants a = 3.890 Å and c = 3.877 Å. **c**, STEM image of the Mn$_3$NiN film with Pt deposited on top of it, interfacial region and the MgO substrate. **d, e**, Strain maps of the $\varepsilon_{xx}$ and $\varepsilon_{yy}$ components of the strain tensor revealing a network of dislocations within the film. **f**, $\varepsilon_{xx}$ and $\varepsilon_{yy}$ averaged over rows of atoms. **g**, FCW and ZFC measurements of magnetic moment in applied magnetic field $\mu_0 H_{ext} = 50\,\text{mT}$ in the IP direction. The vertical arrow indicates the Néel temperature. **h**, Spectrum of optical transmission of the sample measured at room temperature.

To characterize the film from structural point of view, we performed XRD measurements using the Malvern Panalytical Empyrean diffractometer, as well as the Strain Field mapping. XRD scan (see Fig. S3a) shows a presence of Laue oscillations around the Mn$_3$NiN (002) peak, not present in the as-grown film, indicating a high degree of crystallinity. This is further confirmed by the RSM around the 113 peak (see Fig. S3b), from which we also extract the lattice constants $c = 3.877$ Å and $a = 3.890$ Å. The unit cell shows a net tetragonality $c/a = 0.997$, indicating a weaker net strain compared to the as-grown film.

Results of the Strain Field Mapping are shown in Fig. S3c-f. We again see Mn$_3$NiN film, a 3 nm thin mixed-phase interface and MgO substrate, and in this case, also Pt layer. Just like for the as-grown film, this layer deposited on top helps the ion milling process. The density of resulting strain fields in both the IP and OOP direction, as well as their amplitudes, is lower than in the as-grown film, further confirming the higher crystal quality.

ZFC and FCW magnetometry measurements (see Fig. S3g) show that both the magnetic moment and $T_N = 242$ K are lower compared to the as-grown film, in agreement with the weaker net strain.

Spectrum of optical transmission (see Fig. S3h) shows no absorption or interference peaks, but compared to the as-grown film, it is much more flat.

## 1.3 Mn$_3$GaN film

The 40 nm thick film of Mn$_3$GaN was grown by the Physical Vapour Deposition method from a MnGa target at 350°C under $2.5 \times 10^{-8}$ Torr at 8% N$_2$ partial pressure into an Ar atmosphere with no annealing.

For structural characterization of the film, we performed XRD measurement, the result of which is shown in Fig. S4a. The OOP XRD resembles that of the as-grown Mn$_3$NiN film, indicating similar crystal quality. This is further confirmed by the measurements of optical transmission spectrum (see Fig. S4b). For magnetic characterization, using a vibrating-sample magnetometer, we measured the dependence of film magnetic moment on applied magnetic field at various temperatures (see Fig. S4c). Even at room temperature, the film shows non-zero coercivity, which is a clear signature of magnetic ordering at this temperature.

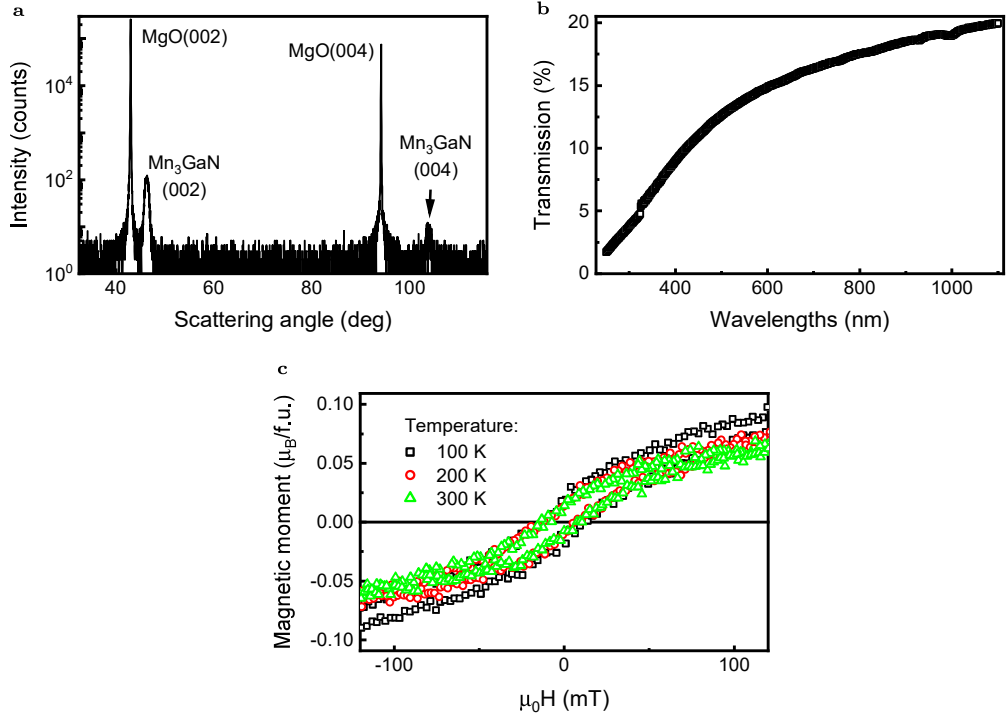

**Fig. S4**: **Mn$_3$GaN sample characterization. a**, XRD scan. **b**, Spectrum of optical transmission of the sample measured at room temperature. **c**, Magnetic moment dependence on applied magnetic field at various temperatures.

# 2 Supplementary Note 2. Additional time-resolved data

## 2.1 Polarization-dependent and -independent parts of the pump-probe signal

Pump-induced change of MO signals in both Mn$_3$NiN and Mn$_3$GaN compounds has two components: pump-polarization-independent (PPI) and pump-polarization-dependent (PPD). As we show in Fig. 1d in the main text, dependence of MO signal on pump linear polarization $\alpha$ is periodic with period 180° and an offset. The periodic part corresponds to PPD signal, while the offset to the PPI signal. To subtract the PPI signal from the as-measured data, which leaves only the PPD part, we used two equivalent methods.

The first method is based on measuring the MO response with pump linear polarization and then with circular polarization. We verified that within the experimental error, opposite helicities of exciting pump pulses generate the same MO signals in both materials. This shows that no helicity-dependent effect are present in these compounds. On top of that, average of MO signals measured with pump linear polarization $\alpha$ and $\alpha + 90°$ is the same as MO signals measured with circularly polarized pump. Therefore, data obtained with a circularly polarized excitation are the PPI part of the signal. This means that subtracting the as-measured data obtained with circular polarization from those obtained with linear polarization leaves only the PPD part of the signal.

The second method is based on measuring the MO response with multiple linear polarizations in the range of $\alpha \in [0° + C, 180° + C)$ (see Fig. S5a where only a subset of the data, for clarity, is shown), with $C$ being an arbitrary angle. Averaging such a set of data leads to PPI signal, as the PPD parts averages to zero. The PPI signal is then subtracted from the as-measured data for a concrete $\alpha$, leaving only the PPD part. Thanks to experimentally measured dependence on linear polarization in these materials, simplified version of this method yielding the PPI signal is to measure only two pump linear polarizations separated by an angle 90° and averaging the signals.

In Fig. S5b we are showing the PPI part obtained using these two methods for two different probe polarizations $\beta$. The data are virtually the same from the time delay $\Delta t \approx 1.5$ ps, confirming the equivalence of these two methods.

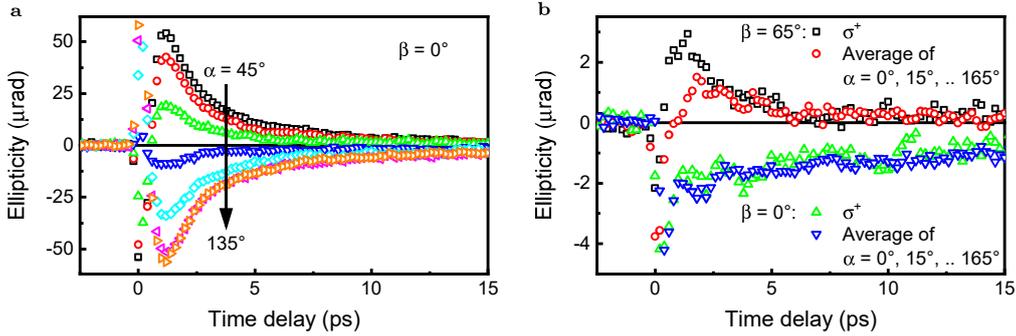

Fig. S5: **Pump-polarization-independent part of the signal.** **a**, As-measured dynamics of ellipticity induced by pump pulses with various linear orientations $\alpha$. Measured with $\beta = 0°$. **b**, Comparing dynamics of ellipticity induced by pump pulses with left circular polarization and an average of dynamics over linear polarization orientations. As-grown Mn$_3$NiN film, $T = 25$ K, $AOI_{probe} \approx 0°$, $I \approx 3.8$ mJ/cm$^2$, $\lambda_{pump} = 870$ nm, $\lambda_{probe} = 435$ nm.

## 2.2 Degree of linear polarization dependence

The methods described in the previous sections allow us to extract the PPD part from the as-measured data. However, it is important to know the relative amplitudes of PPD and PPI signals. The quantity *degree of linear polarization dependence*

$$DLPD(\alpha) \equiv 100\% \frac{|\Delta\gamma_\alpha|}{|\Delta\gamma_\alpha| + |\Delta\gamma_\sigma|} \tag{S2.1}$$

serves as a good measure. $\Delta\gamma_\alpha$ represents the as-measured change of ellipticity for pump linear polarization $\alpha$ and $\Delta\gamma_\sigma$ represents the PPI signal. Equivalently, this quantity can be defined for the pump-induced change of rotation $\Delta\beta$.

Assuming only positive (or only negative) measured values of ellipticity change, $DLPD \approx 100\%$ means that the as-measured MO signal induced with pump linear polarization is much larger than that with pump circular polarization and $DLPD \approx 50\%$ represents similar strengths of the signals. Similarly, $DLPD \approx 100\%$ means PPD $>>$ PPI, $DLPD \approx 67\%$ means PPD $\approx$ PPI, while $DLPD \approx 50\%$ means PPD $<<$ PPI. This comes from the facts that $\Delta\gamma_\alpha = \Delta\gamma_{\text{PPI}} + \Delta\gamma_{\text{PPD}}$ and $\Delta\gamma_\sigma = \Delta\gamma_{\text{PPI}}$.

Actual experimental values that we measure could be both positive and negative, even zero. In case when the PPD and PPI parts have comparable magnitudes and opposite signs, the quantity $DLPD$ does not have a good physical meaning. For example, if the PPD part of the signal is $50\,\mu$rad and PPI is $-50\,\mu$rad, it results in $\Delta\gamma_\alpha = \text{PPI} + \text{PPD} = 0$, and therefore $DLPD = 0$.

$DLPD$ is dependent on multiple parameters. Apart from the obvious dependence on orientation of pump linear polarization $\alpha$, it can also depend on the time delay $\Delta t$ between pump and probe pulses, on the orientation of probe polarization $\beta$, angle of incidence of probe $AOI_{probe}$, etc.

Fig. S6a,b show how $DLDP$ depends on $\alpha$ for $\beta = 0°$ and $65°$ in Mn$_3$NiN. As one could expect from Fig. 1b-d in the main text, $DLPD$ is largest for the cases when the difference between $\alpha$ and $\beta$ is close to $45°$ or $135°$ and smallest when the difference is close to $0°$ or $90°$. The curves displayed in Fig. 1e in the main text are averages of the $DLPD(\alpha = 135°)$ and $DLPD(\alpha = 45°)$ for $\beta = 0°$ and $DLPD(\alpha = 105°)$ and $DLPD(\alpha = 15°)$ for $\beta = 65°$.

Fig. S6c shows how $DLDP$ depends on $\alpha$ for $\beta = 135°$ in Mn$_3$GaN. It has a similar dependence as in Mn$_3$NiN, but due to larger absolute as-measured signals (see Fig. 3 in the main text), $DLPD$ in Mn$_3$GaN is less noisy.

As we can see, in both materials we are reaching values of $DLPD$ close to 100%. This shows that under certain conditions, the main contribution to the MO signal is coming from PPD part of the signal, not the PPI. As we will describe in details in Sec. 2.5, the crucial parameter is $AOI_{probe}$. The closer it is to the normal incidence, the larger the value of $DLPD$ is.

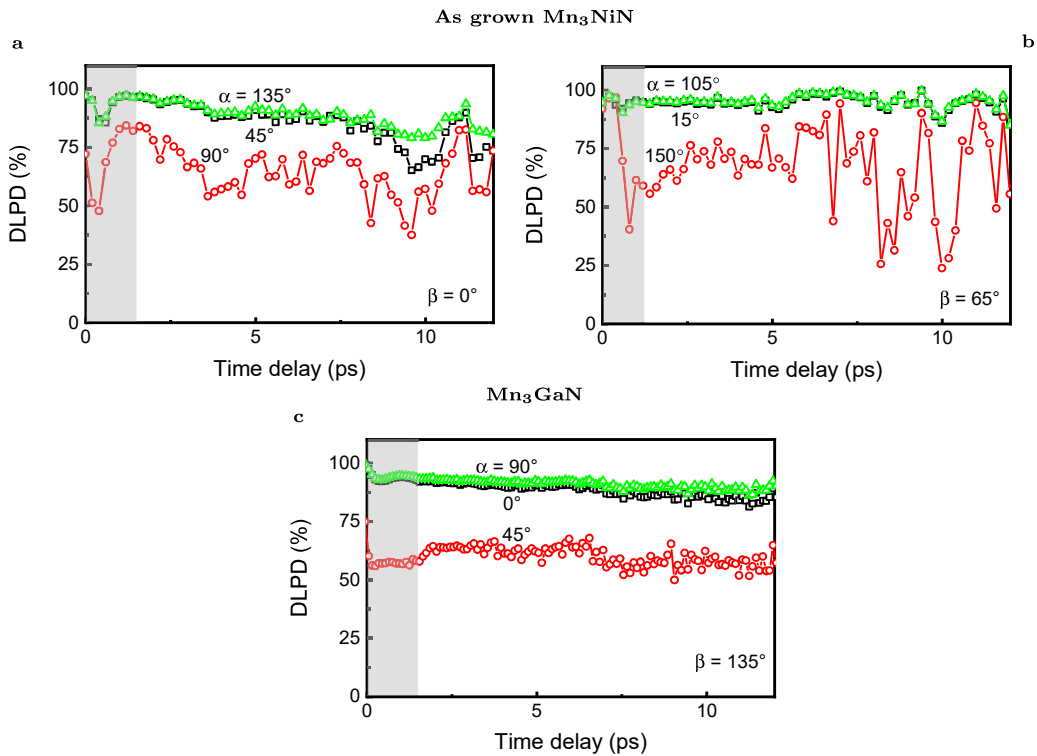

**Fig. S6**: **Degree of linear polarization dependence. a,b**, DLPD in as-grown Mn$_3$NiN film for various pump polarization orientations $\alpha$ and two probe polarizations $\beta$. $T = 25$ K, $AOI_{probe} \approx 0°$, $I \approx 3.8\,\text{mJ}/\text{cm}^2$, $\lambda_{pump} = 870$ nm, $\lambda_{probe} = 435$ nm. **c**, DLPD in Mn$_3$GaN film for various $\alpha$ and $\beta = 135°$. $T = 25$ K, $AOI_{probe} \approx 0°$, $I \approx 5.2\,\text{mJ}/\text{cm}^2$, $\lambda_{pump} = 820$ nm, $\lambda_{probe} = 532$ nm.

## 2.3 Probe polarization dependence

Fig. 2c in the main text shows that for a constant $\alpha$, the detected MO signal depends harmonically on the input probe polarization $\beta$. This is a fingerprint of the Voigt effect (sometimes termed as the Cotton-Mouton effect), typically used to detect magnetic changes in antiferromagnets, as it is an effect quadratic in magnetization [12].

In Fig. S7a we plot the PPD part of the ellipticity change detected with various $\beta$. As expected, the largest signals correspond to the cases when the difference between $\alpha$ and $\beta$ is either $45°$ or $135°$. Fig. S7b, where the dependence of ellipticity change at time delay $\Delta t = 3$ ps on $\beta$ for multiple values of $\alpha$ is plotted, shows the expected harmonic dependence. In accord with Fig. 1d in the main text, we see a phase shift of the sinusoidal dependence for different values of $\alpha$ and this shift directly corresponds to the value of $\alpha$. Moreover, we again see that the maximum value of ellipticity change is independent of pump polarization orientation $\alpha$.

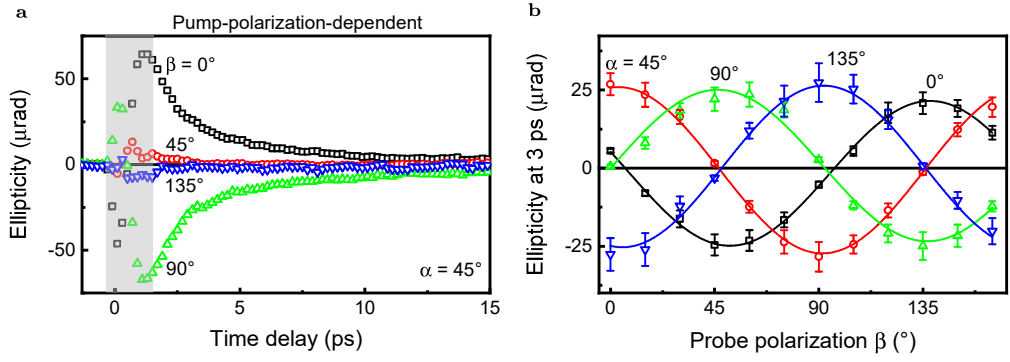

**Fig. S7**: **Probe-polarization dependence. a**, Dynamics of the PPD part of ellipticity change measured with various $\beta$ induced by pump pulses with $\alpha = 45°$. **b**, Probe-polarization dependence of ellipticity change for $\Delta t = 3\,\text{ps}$ induced by pump pulses with four different $\alpha$. As-grown Mn$_3$NiN film, $T = 25\,\text{K}$, $AOI_{probe} \approx 0°$, $I \approx 3.8\,\text{mJ/cm}^2$, $\lambda_{pump} = 870\,\text{nm}$, $\lambda_{probe} = 435\,\text{nm}$.

## 2.4 Transient transmission

Pump-induced change of sample transmittance, $\Delta T/T$, known as differential or transient transmission, is a quantity often used to characterise the dynamics of electrons in metallic systems [13–16]. For example, it could describe the flow of energy between electrons and phonons, the thermalization time of electron bath, the rise of electron temperature, etc. In Fig. S8a,b we show the dynamics of $\Delta T/T$ in Mn$_3$NiN for various pump polarization states in 15 ps and 3.5 ns time scales. All the curves overlap, which means that the energy absorbed from the pump pulse by the system is the same independent on its polarization. Moreover, the longest time constant reflects the heat dissipation from the Mn$_3$NiN film to the MgO substrate [17].

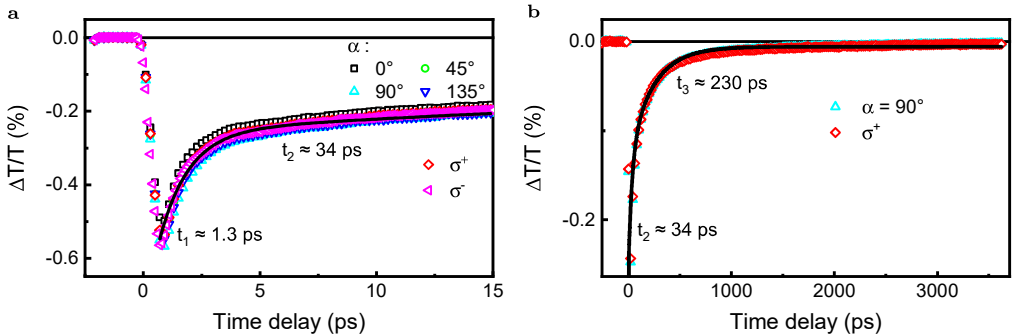

**Fig. S8**: **Dynamics of transient transmission. a**, 15 ps time scale for dynamics of $\Delta T/T$ induced by pump pulses with various $\alpha$. **b**, As in **a**, but 3.5 ns time scale. The solid black line in both graphs represents a three-exponential fit of data shown in **a** and **b** with decay times of 1.3 ps, 34 ps and 230 ps, sign amplitudes -0.51 %, -0.13 % and -0.13 %, respectively and an offset -0.006 %. As-grown Mn$_3$NiN film, $T = 25\,\text{K}$, $AOI_{probe} \approx 0°$, $I \approx 3.8\,\text{mJ/cm}^2$, $\lambda_{pump} = 870\,\text{nm}$, $\lambda_{probe} = 435\,\text{nm}$.

## 2.5 Angle of incidence

In our experimental setup, the angle between the pump and probe beams was fixed to $\approx 15°$, but the sample could be oriented at different angles with respect to the incident laser beams. In Fig. S9a we are showing two typical cases of sample rotations: 1$^{st}$ configuration - normal incidence of probe beam, angle of pump beam incidence being around $-15°$; 2$^{nd}$ configuration - normal incidence of pump beam, angle of probe beam incidence being around 15°.

Fig. S9b,c show the dynamics of the PPD and PPI signals, respectively, for the two aforementioned cases of angles of incidences, i.e., sample rotations. PPD signals are virtually independent of sample rotation, which is in agreement that the detection mechanism is the Voigt effect, while PPI signals change drastically. We observed that, in general, the closer the probe beam is to the normal incidence, the weaker the PPI part of the signal is. This is valid for both Mn$_3$NiN and Mn$_3$GaN. This PPI signal comes from MO effects that are linear in magnetic moment, as will be discussed in our following publication [18].

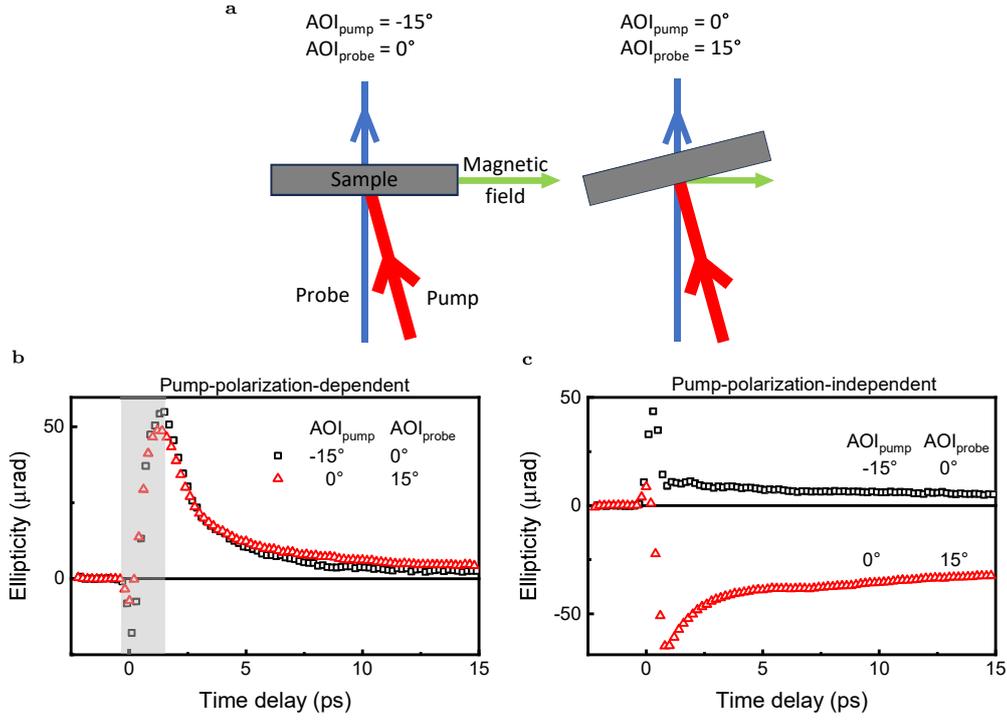

**Fig. S9**: **Angle of incidence. a**, Schematics of the geometry of the pump-probe set-up. Rotation of the sample changes angle of incidence of both pump and probe beams. **b**, Dynamics of pump-polarization-dependent part of the ellipticity for the two cases of sample rotation shown in **a**. **c**, As in **b**, but dynamics of pump-polarization-independent part of the ellipticity is shown. As-grown $Mn_3NiN$ film, $T = 25\,\text{K}$, $I \approx 3.8\,\text{mJ/cm}^2$, $\lambda_{pump} = 870\,\text{nm}$, $\lambda_{probe} = 435\,\text{nm}$, $\alpha = 0°$, $\beta = 45°$.

## 2.6 Fluence of the pump

To gain more insight into the behaviour of PPD and PPI signals, we systematically studied their dynamics for various pump fluences. Since the PPI part is suppressed for the probe normal incidence, we used the geometry for normal incidence of pump, i.e., $AOI_{probe} \approx 15°$, for which the PPI signal is present, too. As-measured data for pump with linear and circular polarizations at various pump fluences are shown in Fig. S10a,b, respectively. In Fig. S10c we show the PPD part of the signal extracted using the first method described in Sec. 2.1. Fig. S10d, shows the dependence of PPD signals at various times delays on the pump fluence. The dynamics of transient transmission $\Delta T/T$ is shown in Fig. S10e.

The dynamics of the PPI (Fig. S10b) part is very similar to that of $\Delta T/T$ (Fig. S10e). This suggests that the PPI part of the MO signal is connected with the pump-induced heating of the studied film, i.e, that this MO signal reflects the demagnetization of the sample by pump pulses [18].

The dynamics of the PPD part is shown in Fig. S10c. Increasing the fluence up to $\approx 5$ mJ/cm$^2$ increases the signal at all time delays. However, further fluence increase leads to a more complicated behaviour. This is clearly visible in Fig. S10d, where we plot the MO signal at various time delays together with a linear dependence, which was obtained by fitting these dependences for small fluences. The observed intensity dependence of MO signal seems to be a super-linear, especially for short time delays. Nevertheless, it is rather difficult to draw any quantitative conclusions from these data because for high intensities the sample temperature increases, which is slowing down the dynamics (see Fig. 2a in the main paper).

The obtained values of DLPD (Fig. S10f,g) are obscured by the opposite signs of PPD and PPI signal parts, as discussed in Chapter 2.2. Nevertheless, Fig. S10g reveals that DLPD value is not significantly affected by the pump fluence.

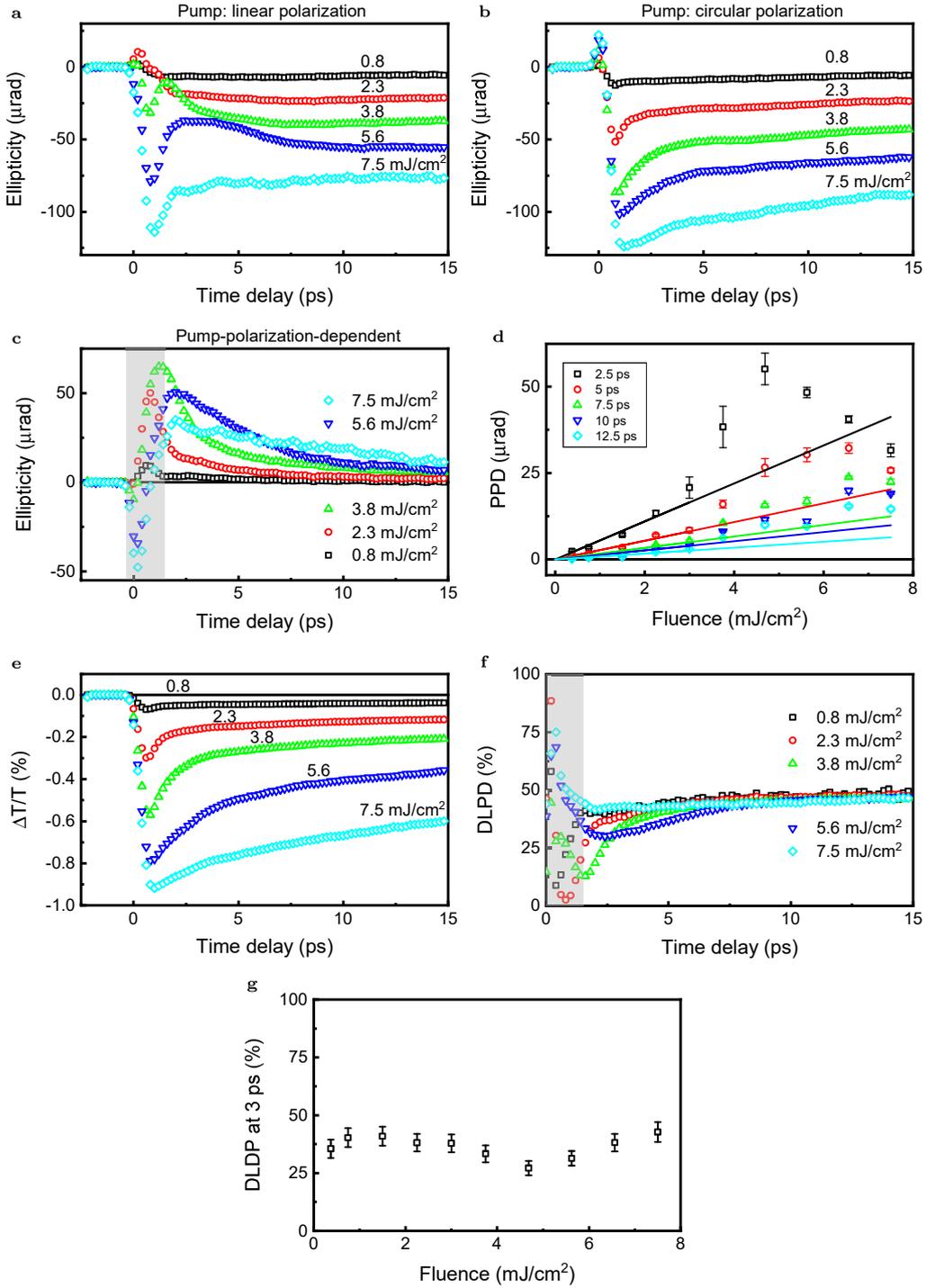

**Fig. S10**: **Dependence on pump fluence.** **a**, As-measured data of dynamics of ellipticity induced by pump pulses with $\alpha = 90°$ and various fluences. **b**, As in **a**, but pump pulses were left-circularly polarized. **c**, Pump-polarization-dependent part of the ellipticity dynamics evaluated from **a** and **b**. **d**, PPD signal as a function of pump fluence for various time delays. Lines are linear fits of fluences $\leq 3\,\text{mJ/cm}^2$ with slopes 5.5, 2.7, 1.7, 1.3 and 0.9 $\mu\text{rad/mJ/cm}^2$. **e**, Dynamics of transient transmission induced by pump pulses with various fluences. **f**, DLPD evaluated from **a** and **b**. **g**, DLDP as a function of pump fluence for $\Delta t = 3\,\text{ps}$. As-grown Mn$_3$NiN film, $T = 25\,\text{K}$, $AOI_{probe} \approx 15°$, $\lambda_{pump} = 870\,\text{nm}$, $\lambda_{probe} = 435\,\text{nm}$, $\beta = 45°$.

## 2.7 Magnetic field independence

Our experimental setup enables to apply external magnetic field in the direction perpendicular to the probe beam and within the plane of incidence of pump beam (see Fig. S9a).

We observed (see Fig. S11) that in the case of probe normal incidence, neither the as-measured change of ellipticity nor the PPD signal nor transient transmission depend on the magnetic field applied during the measurements within the range $\mu_0 H = \pm 530\,\mathrm{mT}$.

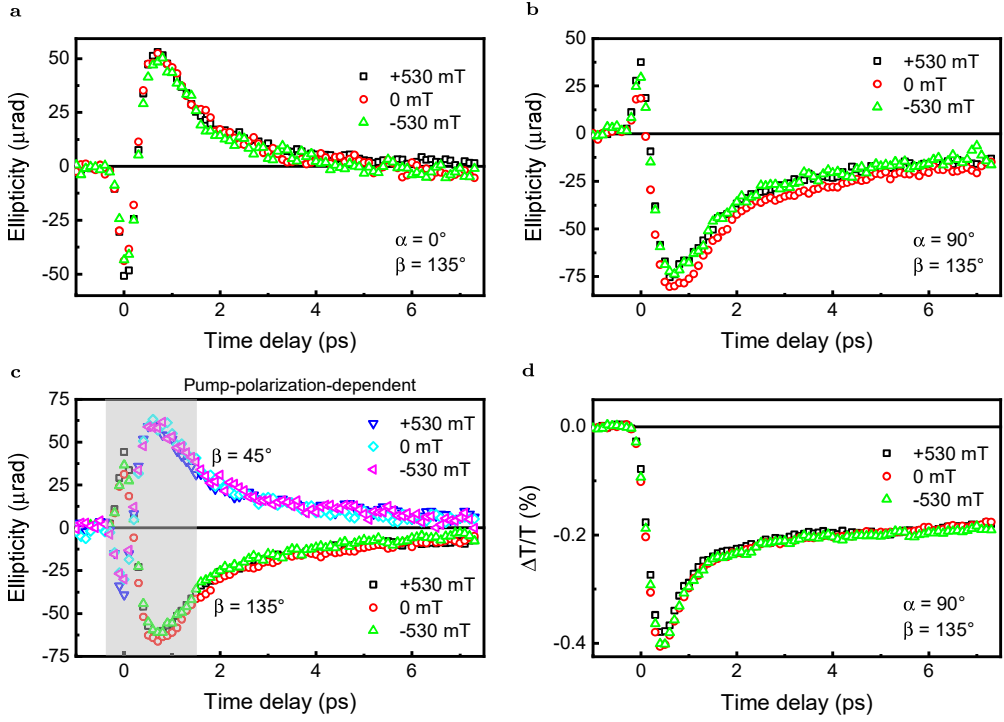

**Fig. S11**: **Independence of MO signal on external magnetic field.** **a**, As-measured data of dynamics of ellipticity while magnetic field was applied in the IP direction. $\alpha = 0°$, $\beta = 135°$. **b**, As in **a**, but $\alpha = 90°$. **c**, PPD parts of the ellipticity dynamics for $\beta = 45°$ and $135°$. **d**, Dynamics of transient transmission with various magnetic field applied. $\alpha = 90°$, $\beta = 135°$. As-grown Mn$_3$NiN film, $T = 25\,\mathrm{K}$, $AOI_{probe} \approx 0°$, $\lambda_{pump} = 800\,\mathrm{nm}$, $\lambda_{probe} = 400\,\mathrm{nm}$, $I \approx 3.8\,\mathrm{mJ/cm^2}$.

We also applied magnetic fields $\mu_0 H = \pm 530\,\mathrm{mT}$ during cooling the sample from the room temperature to $T = 25\,\mathrm{K}$. However, no difference is seen in the PPD signal (see Fig. S12).

Observed independences are not surprising for two reasons. First, since the coercive field of our Mn$_3$NiN sample could be as large as $\approx 1.5\,\mathrm{T}$ at temperature $T = 25\,\mathrm{K}$ (see Fig. S2c), one could expect that a magnetic field with strength $\approx 0.53\,\mathrm{T}$ would not have a strong effect. Second, our detailed analysis of Mn$_3$NiN samples in [4] shows that regions with opposite domain types not responding to an applied field are commonly present. Their presence is linked with dislocations, which we did observe in our samples, too (see Fig. S1c-f).

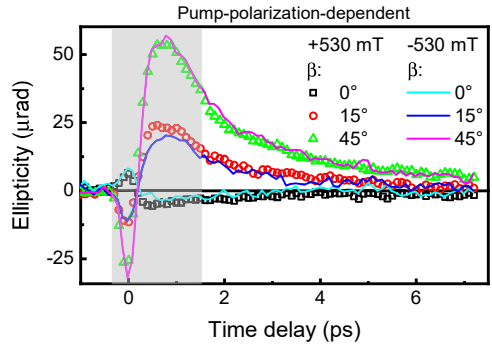

**Fig. S12**: **Independence of MO signal on field cooling.** Pump-polarization-dependent part of the ellipticity change measured with various $\beta$ for two opposite directions of magnetic field. The field $\mu_0 H_{ext} = \pm 530\,\mathrm{mT}$ was applied in the IP direction from $T = 300\,\mathrm{K}$ to $25\,\mathrm{K}$, where the data was measured. As-grown Mn$_3$NiN film, $AOI_{probe} \approx 0°$, $\lambda_{pump} = 800\,\mathrm{nm}$, $\lambda_{probe} = 400\,\mathrm{nm}$, $I \approx 3.8\,\mathrm{mJ/cm^2}$.

## 2.8 Magnetic origin of the PPD signals

Generally, the experimentally measured MO signals are sensitive not only to the magnetic dynamics of studied system, but also to a pump-induced change of the complex index of refraction. This signal is often referred to as the *optical part* of the MO signal. In order to verify if it is the *magnetic* or the *optical* part that dominates in the measured data, both pump-induced change of rotation $\Delta\beta$ and ellipticity $\Delta\gamma$ should be measured and their dynamics need to be compared [19].

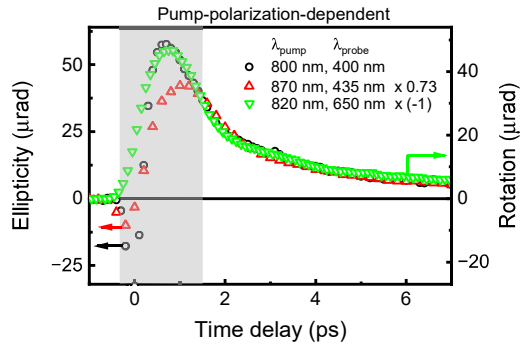

Fig. S13: **Pump-induced dynamics of ellipticity and rotation at various wavelengths.** Dynamics of the pump-polarization-dependent part of the ellipticity in Mn$_3$NiN for $\lambda_{pump} = 800$ nm, $\lambda_{probe} = 400$ nm (black circles), $\lambda_{pump} = 870$ nm, $\lambda_{probe} = 435$ nm (red up triangles) and of rotation for $\lambda_{pump} = 820$ nm, $\lambda_{probe} = 650$ nm (green down triangles). $T = 25$ K, $I \approx 3.8$ mJ/cm$^2$, $\alpha = 90°$, $\beta = 45°$, $AOI_{probe} \approx 0°$ for ellipticity, $AOI_{probe} \approx 15°$ for rotation.

In the static case when only the Voigt effect is present, rotation $\beta$ and ellipticity $\gamma$ could be described as

$$\beta = f_\beta M^2, \tag{S2.2}$$

$$\gamma = f_\gamma M^2, \tag{S2.3}$$

where $f_\beta$ and $f_\gamma$ are optical "constant" dependent on material electronic properties and light wavelength, and $M$ is magnetization.

If a pump pulse induces a material change to the studied material, it leads to a change of polarization rotation $\Delta\beta(t)$ and ellipticity $\Delta\gamma(t)$. Since both *optical* and *magnetic* parts could be modified by the pump pulse, we have

$$\Delta\beta(t) \approx \Delta f_\beta(t) M^2 + f_\beta\, 2M\, \Delta M(t), \tag{S2.4}$$

$$\Delta\gamma(t) \approx \Delta f_\gamma(t) M^2 + f_\gamma\, 2M\, \Delta M(t). \tag{S2.5}$$

The first terms on the right-hand sides are the *optical* parts, the second terms are the *magnetic* parts. One can see that in general case, the dynamics of rotation and ellipticity does not need to be same. However, if the *magnetic* parts dominate, i.e., if $|\Delta f_\beta M^2| \ll |f_\beta\, 2M\, \Delta M|$ and $|\Delta f_\gamma M^2| \ll |f_\gamma\, 2M\, \Delta M|$, then

$$\Delta\beta(t) \approx \frac{f_\beta}{f_\gamma} \Delta\gamma(t) \approx f_\beta\, 2M\, \Delta M(t), \tag{S2.6}$$

which means that up to a multiplicative constant, the dynamics of rotation and ellipticity must be equal, and that $\Delta\beta(t)$, $\Delta\gamma(t)$ represent magnetic dynamics. Similar analysis shows that measurements of dynamics of rotation (or of ellipticity) for various probe wavelengths must lead to the same dynamics, up to a multiplicative constant.

We measured that the dynamics of PPD part of rotation and ellipticity is the same after a certain time delay in both Mn$_3$NiN and Mn$_3$GaN films (Fig. 2b and 3b in the main text). Moreover, we show in Fig. S13 that this holds also for different probe wavelengths. Finally, PPD signals are pronounced only below $T_N$ and (almost) zero above it (Fig. 2a and 3e in the main text). This shows that we detect, after the transient optical effects are gone (i.e., after $\approx 1.5$ ps in Mn$_3$NiN and $\approx 500$ fs in Mn$_3$GaN), pump-induced changes of magnetic ordering.

## 2.9 Annealed Mn$_3$NiN sample

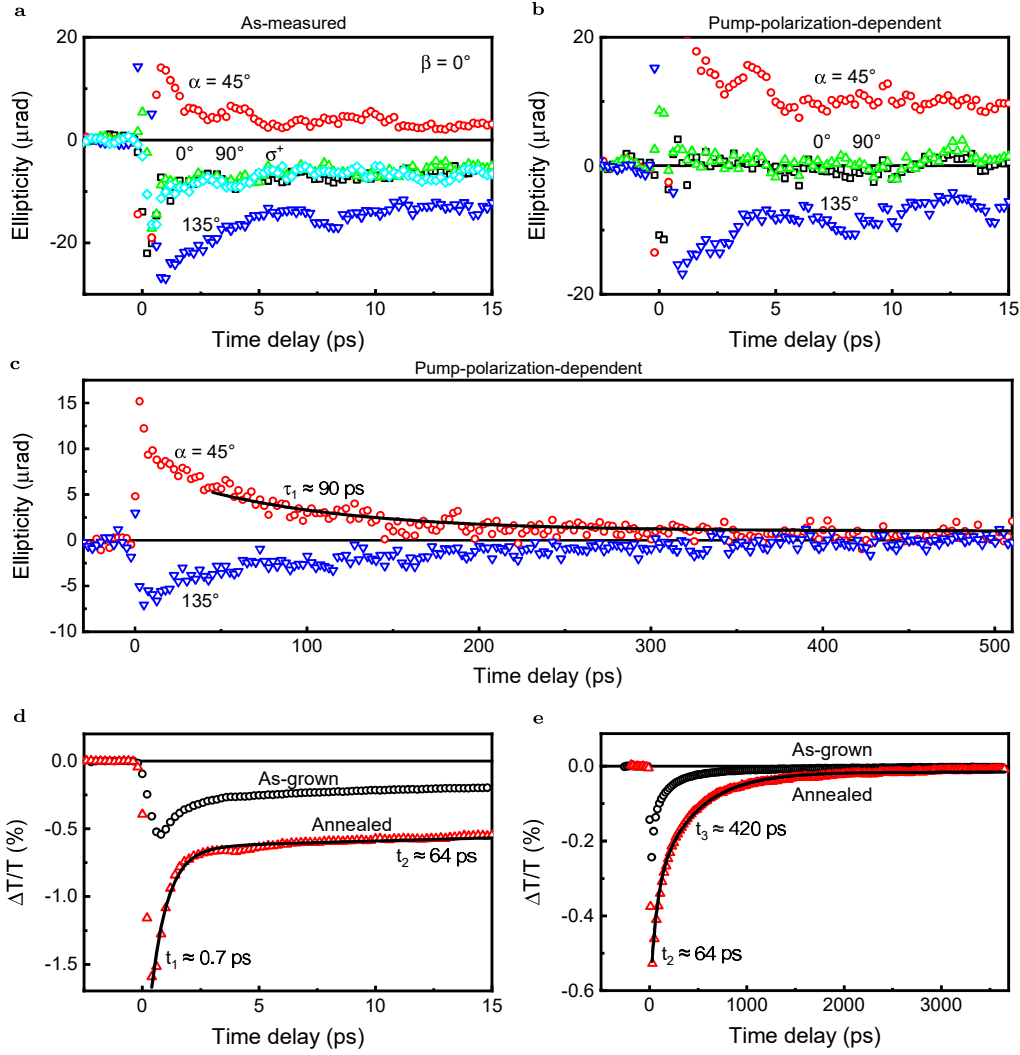

**Fig. S14**: **Magnetic dynamics in Mn$_3$NiN annealed sample.** **a**, As-measured data of ellipticity dynamics in annealed sample of Mn$_3$NiN induced by pump pulses with various polarization states. **b**, As in **a**, but only pump-polarization-dependent part is shown. **c**, 500-ps time scale for dynamics of ellipticity in annealed sample for two linear polarizations of pump pulses. The solid line is a monoexponential fit. **d**, Comparison of transient transmission dynamics of as-grown and annealed samples of Mn$_3$NiN. **e**, As in **d**, but at longer time scale. The solid line in **d** and **e** represents a three-exponential fit of data for the annealed sample with decay times of 0.7 ps, 64 ps and 420 ps, sign amplitudes -1.8 %, -0.29 % and -0.32 %, respectively and an offset -0.016 %. $T = 25$ K, $I \approx 3.8$ mJ/cm$^2$, $AOI_{probe} \approx 0°$, $\lambda_{pump} = 870$ nm, $\lambda_{probe} = 435$ nm, $\beta = 0°$.

In order to experimentally verify whether the presence of local dislocations in the Mn$_3$NiN film (see Fig. S1c-f) is responsible for the observed independence of the PPD signals with respect to the crystallographic orientation in our epitaxial films, we measured the pump-induced change of ellipticity also in the annealed sample (Sec. 1.2), where their density is considerably lower (see Fig. S3c-f). In Fig. S14a, b we show the as-measured and PPD data for different pump polarization states, respectively. One could see that, even though the signals are smaller than for the as-grown sample, the symmetry is the same, i.e., the PPD signal is the strongest when the difference between $\alpha$ and $\beta$ is $\approx 45°$. On the other hand, the signals do not decay as fast as for the as-grown film, where the decay time was $\tau_1 \approx 2.3$ ps (see Fig. 1b in the main text). Measuring the PPD signal in the annealed sample within 500 ps time delay range (Fig. S14c) shows a much slower decay, with $\tau_1 \approx 90$ ps. This suggests that the density of local dislocations do not play a major role in the mechanism responsible for the signal generation. On the other hand, the crystallographic quality of the film does change the dynamics of the PPD signal - the measured decay of the PPD part of the MO signal is considerably slower in the film with the higher crystallographic quality.

Fig. S14d,e show how the dynamics of transient transmission differs in the annealed and as-grown samples. The absolute amplitude is larger and the signal decays for a longer time in the annealed sample.

## 2.10 Mn$_3$GaN at room temperature

The PPD signal in Mn$_3$GaN decreases rapidly with increasing temperature above 200 K (Fig. 3e in the main text). Nevertheless, even at $T = 296$ K the PPD signal exists, see Fig. S15. This shows that Mn$_3$GaN is a material where the pump-polarization sensitivity exists even at room temperature.

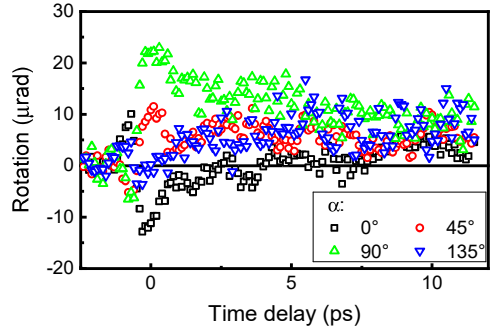

**Fig. S15**: **Mn$_3$GaN at room temperature.** As-measured dynamics of rotation induced by pump pulses with 4 different linear pump polarizations. $\lambda_{pump} = 820$ nm, $\lambda_{probe} = 532$ nm. $T = 296$ K, $I \approx 5.2$ mJ/cm$^2$, $\beta = 135°$, $AOI_{probe} \approx 0°$.

## 2.11 Fluence dependence in Mn$_3$GaN

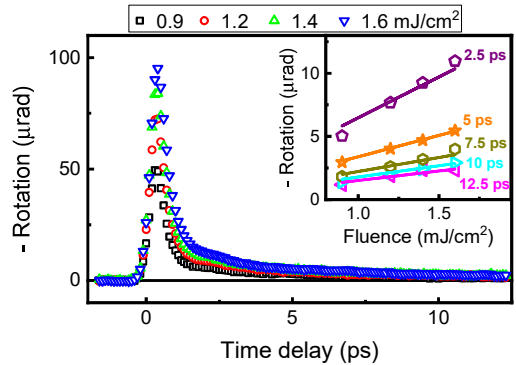

**Fig. S16**: **Fluence dependence in Mn$_3$GaN.** As-measured dynamics of rotation induced by pump pulses with various fluences. Inset: Fluence dependence at various time delays. Lines are linear fits with slopes 1.5, 1.8, 2.2, 3.4 and 6.5 $\mu$rad/mJ/cm$^2$. $\lambda_{pump} = 820$ nm, $\lambda_{probe} = 532$ nm. $T = 100$ K, $\alpha = 0°$, $\beta = 135°$, $AOI_{probe} \approx 0°$.

Similarly to Mn$_3$NiN, we studied the dependence of MO signals on pump fluence in Mn$_3$GaN, too. In Fig. S15 we show the intensity dependence of the as-measured dynamics of rotation. As apparent in the inset, in the investigated intensity range the dynamics is linearly proportional to the pump fluence.

# 3 Supplementary Note 3. Symmetry analysis

To describe the symmetry of the second-order torque, we use the microscopic formula derived in [20], which allows describing the torque for both circular and linearly polarized light. In this approach, the response is described by a complex response tensor $\chi_{ijk}$ such that the torque is

$$T_i = \mathrm{Im}\left[\chi_{ijk} e_j e_k^*\right]. \tag{S3.1}$$

Here $\mathbf{e}$ is the polarization vector defined such that the electric field is given by $\mathbf{E}(t) = \mathrm{Re}\left[E_0 \mathbf{e} \exp(-i\omega t)\right]$, where $E_0$ is the amplitude of the electric field and $\omega$ is the frequency. Note that in Eq. (S3.1) all the prefactors have been included in the response tensor $\chi$ as these do not matter for symmetry.

By swapping the indices $j$ and $k$ in Eq. (S3.2) we find:

$$T_i = \mathrm{Im}\left[\chi_{ijk} \epsilon_j \epsilon_k^*\right] = -\mathrm{Im}\left[\chi_{ikj}^* \epsilon_j \epsilon_k^*\right]. \tag{S3.2}$$

Since this has to hold for all $\mathbf{e}$ we have additional constraint on the tensor $\chi$:

$$\chi_{ijk} = -\chi_{ikj}^*. \tag{S3.3}$$

Apart from this constraint, the symmetrization of the tensor $\chi$ can be done using standard approach based on the Neumann's principle. This tensor can be separated into time-reversal odd ($\mathcal{T}$-odd) and time-reversal even ($\mathcal{T}$-even) components, as is commonly done for linear response effects. Formally, this can be defined as

$$\chi^{\mathrm{odd}} = (\chi - \chi^{\mathcal{T}})/2, \tag{S3.4}$$
$$\chi^{\mathrm{even}} = (\chi + \chi^{\mathcal{T}})/2, \tag{S3.5}$$

where $\chi^{\mathcal{T}}$ is the response tensor for the time-reversed system.

We have done all the symmetry analysis using an open-source Symmetr code [21]. From technical point of view the symmetrization is done exactly as for normal real response tensor since the symmetry operations are represented by real matrices, the only difference is the additional constraint (S3.3).

The torque can be represented by an effective magnetic field such that $\mathbf{T}_n = \mathbf{M}_n \times \mathbf{B}_n$, where $n = 1, 2$ or 3 denotes the sublattice. Note that the effective field corresponding to the $\mathcal{T}$-even torque is $\mathcal{T}$-odd and vice-versa. To keep the notation simple we label the effective fields according to the torque transformation:

$$\mathbf{T}_n^{\mathrm{odd}} = \mathbf{M}_n \times \mathbf{B}_n^o, \tag{S3.6}$$
$$\mathbf{T}_n^{\mathrm{even}} = \mathbf{M}_n \times \mathbf{B}_n^e. \tag{S3.7}$$

The symmetry-restricted forms for the effective fields corresponding to the $\mathcal{T}$-odd torque induced by linearly-polarized light in the absence of spin-orbit coupling are given in Table S3.1. We give the results for all 8 domains of Mn$_3$NiN (see Fig. S18). The analogous results for the $\mathcal{T}$-even torque are given in Table S3.2. $B^o$, $B^e$ are parameters that express the strength of the effective fields. Note that here we give only the component of the effective field that is perpendicular to the magnetic moment since only this component matters for the torque.

The results for circularly polarized light are given in Tables S3.3 and S3.4. We find that in contrast to ferromagnets, the results are the same for left and right-polarized light.

| Domain # | $B_1$ | $B_2$ | $B_3$ |
|---|---|---|---|
| 1 | $-B^o \sin^2(\alpha)\begin{bmatrix}1 & -1 & 1\end{bmatrix}$ | $B^o \cos^2(\alpha)\begin{bmatrix}1 & -1 & 1\end{bmatrix}$ | $-B^o \cos(2\alpha)\begin{bmatrix}1 & -1 & 1\end{bmatrix}$ |
| 2 | $B^o \sin^2(\alpha)\begin{bmatrix}1 & 1 & 1\end{bmatrix}$ | $-B^o \cos^2(\alpha)\begin{bmatrix}1 & 1 & 1\end{bmatrix}$ | $B^o \cos(2\alpha)\begin{bmatrix}1 & 1 & 1\end{bmatrix}$ |
| 3 | $B^o \sin^2(\alpha)\begin{bmatrix}1 & -1 & -1\end{bmatrix}$ | $-B^o \cos^2(\alpha)\begin{bmatrix}1 & -1 & -1\end{bmatrix}$ | $B^o \cos(2\alpha)\begin{bmatrix}1 & -1 & -1\end{bmatrix}$ |
| 4 | $-B^o \sin^2(\alpha)\begin{bmatrix}1 & 1 & -1\end{bmatrix}$ | $B^o \cos^2(\alpha)\begin{bmatrix}1 & 1 & -1\end{bmatrix}$ | $-B^o \cos(2\alpha)\begin{bmatrix}1 & 1 & -1\end{bmatrix}$ |
| 5 | $-B^o \sin^2(\alpha)\begin{bmatrix}1 & -1 & 1\end{bmatrix}$ | $B^o \cos^2(\alpha)\begin{bmatrix}1 & -1 & 1\end{bmatrix}$ | $-B^o \cos(2\alpha)\begin{bmatrix}1 & -1 & 1\end{bmatrix}$ |
| 6 | $B^o \sin^2(\alpha)\begin{bmatrix}1 & 1 & 1\end{bmatrix}$ | $-B^o \cos^2(\alpha)\begin{bmatrix}1 & 1 & 1\end{bmatrix}$ | $B^o \cos(2\alpha)\begin{bmatrix}1 & 1 & 1\end{bmatrix}$ |
| 7 | $B^o \sin^2(\alpha)\begin{bmatrix}1 & -1 & -1\end{bmatrix}$ | $-B^o \cos^2(\alpha)\begin{bmatrix}1 & -1 & -1\end{bmatrix}$ | $B^o \cos(2\alpha)\begin{bmatrix}1 & -1 & -1\end{bmatrix}$ |
| 8 | $-B^o \sin^2(\alpha)\begin{bmatrix}1 & 1 & -1\end{bmatrix}$ | $B^o \cos^2(\alpha)\begin{bmatrix}1 & 1 & -1\end{bmatrix}$ | $-B^o \cos(2\alpha)\begin{bmatrix}1 & 1 & -1\end{bmatrix}$ |

**Table S3.1**: The effective field corresponding to the $\mathcal{T}$-odd torque in absence of spin-orbit coupling for linearly polarized light with the polarization along the [$\cos(\alpha)$,$\sin(\alpha)$,0] direction.

| Domain # | $B_1$ | $B_2$ | $B_3$ |
|---|---|---|---|
| 1 | $B^e \sin^2(\alpha) \begin{bmatrix} 0 & 1 & 1 \end{bmatrix}$ | $-B^e \cos^2(\alpha) \begin{bmatrix} 1 & 0 & -1 \end{bmatrix}$ | $-B^e \cos(2\alpha) \begin{bmatrix} 1 & 1 & 0 \end{bmatrix}$ |
| 2 | $-B^e \sin^2(\alpha) \begin{bmatrix} 0 & 1 & -1 \end{bmatrix}$ | $-B^e \cos^2(\alpha) \begin{bmatrix} 1 & 0 & -1 \end{bmatrix}$ | $-B^e \cos(2\alpha) \begin{bmatrix} 1 & -1 & 0 \end{bmatrix}$ |
| 3 | $-B^e \sin^2(\alpha) \begin{bmatrix} 0 & 1 & -1 \end{bmatrix}$ | $B^e \cos^2(\alpha) \begin{bmatrix} 1 & 0 & 1 \end{bmatrix}$ | $B^e \cos(2\alpha) \begin{bmatrix} 1 & 1 & 0 \end{bmatrix}$ |
| 4 | $B^e \sin^2(\alpha) \begin{bmatrix} 0 & 1 & 1 \end{bmatrix}$ | $B^e \cos^2(\alpha) \begin{bmatrix} 1 & 0 & 1 \end{bmatrix}$ | $B^e \cos(2\alpha) \begin{bmatrix} 1 & -1 & 0 \end{bmatrix}$ |
| 5 | $-B^e \sin^2(\alpha) \begin{bmatrix} 0 & 1 & 1 \end{bmatrix}$ | $B^e \cos^2(\alpha) \begin{bmatrix} 1 & 0 & -1 \end{bmatrix}$ | $B^e \cos(2\alpha) \begin{bmatrix} 1 & 1 & 0 \end{bmatrix}$ |
| 6 | $B^e \sin^2(\alpha) \begin{bmatrix} 0 & 1 & -1 \end{bmatrix}$ | $B^e \cos^2(\alpha) \begin{bmatrix} 1 & 0 & -1 \end{bmatrix}$ | $B^e \cos(2\alpha) \begin{bmatrix} 1 & -1 & 0 \end{bmatrix}$ |
| 7 | $B^e \sin^2(\alpha) \begin{bmatrix} 0 & 1 & -1 \end{bmatrix}$ | $-B^e \cos^2(\alpha) \begin{bmatrix} 1 & 0 & 1 \end{bmatrix}$ | $-B^e \cos(2\alpha) \begin{bmatrix} 1 & 1 & 0 \end{bmatrix}$ |
| 8 | $-B^e \sin^2(\alpha) \begin{bmatrix} 0 & 1 & 1 \end{bmatrix}$ | $-B^e \cos^2(\alpha) \begin{bmatrix} 1 & 0 & 1 \end{bmatrix}$ | $-B^e \cos(2\alpha) \begin{bmatrix} 1 & -1 & 0 \end{bmatrix}$ |

**Table S3.2**: The effective field corresponding to the $\mathcal{T}$-even torque in absence of spin-orbit coupling for linearly polarized light with the polarization along the $[\cos(\alpha),\sin(\alpha),0]$ direction.

| Domain # | $B_1$ | $B_2$ | $B_3$ |
|---|---|---|---|
| 1 | $-B^o \begin{bmatrix} 1 & -1 & 1 \end{bmatrix}$ | $B^o \begin{bmatrix} 1 & -1 & 1 \end{bmatrix}$ | $\begin{bmatrix} 0 & 0 & 0 \end{bmatrix}$ |
| 2 | $B^o \begin{bmatrix} 1 & 1 & 1 \end{bmatrix}$ | $-B^o \begin{bmatrix} 1 & 1 & 1 \end{bmatrix}$ | $\begin{bmatrix} 0 & 0 & 0 \end{bmatrix}$ |
| 3 | $B^o \begin{bmatrix} 1 & -1 & -1 \end{bmatrix}$ | $-B^o \begin{bmatrix} 1 & -1 & -1 \end{bmatrix}$ | $\begin{bmatrix} 0 & 0 & 0 \end{bmatrix}$ |
| 4 | $-B^o \begin{bmatrix} 1 & 1 & -1 \end{bmatrix}$ | $B^o \begin{bmatrix} 1 & 1 & -1 \end{bmatrix}$ | $\begin{bmatrix} 0 & 0 & 0 \end{bmatrix}$ |
| 5 | $-B^o \begin{bmatrix} 1 & -1 & 1 \end{bmatrix}$ | $B^o \begin{bmatrix} 1 & -1 & 1 \end{bmatrix}$ | $\begin{bmatrix} 0 & 0 & 0 \end{bmatrix}$ |
| 6 | $B^o \begin{bmatrix} 1 & 1 & 1 \end{bmatrix}$ | $-B^o \begin{bmatrix} 1 & 1 & 1 \end{bmatrix}$ | $\begin{bmatrix} 0 & 0 & 0 \end{bmatrix}$ |
| 7 | $B^o \begin{bmatrix} 1 & -1 & -1 \end{bmatrix}$ | $-B^o \begin{bmatrix} 1 & -1 & -1 \end{bmatrix}$ | $\begin{bmatrix} 0 & 0 & 0 \end{bmatrix}$ |
| 8 | $-B^o \begin{bmatrix} 1 & 1 & -1 \end{bmatrix}$ | $B^o \begin{bmatrix} 1 & 1 & -1 \end{bmatrix}$ | $\begin{bmatrix} 0 & 0 & 0 \end{bmatrix}$ |

**Table S3.3**: The effective field corresponding to the $\mathcal{T}$-odd torque in absence of spin-orbit coupling for circularly polarized light propagating along the $z$ direction. $\mathbf{e} = [0, i, 0]$.

| Domain # | $B_1$ | $B_2$ | $B_3$ |
|---|---|---|---|
| 1 | $B^e \begin{bmatrix} 0 & 1 & 1 \end{bmatrix}$ | $-B^e \begin{bmatrix} 1 & 0 & -1 \end{bmatrix}$ | $\begin{bmatrix} 0 & 0 & 0 \end{bmatrix}$ |
| 2 | $-B^e \begin{bmatrix} 0 & 1 & -1 \end{bmatrix}$ | $-B^e \begin{bmatrix} 1 & 0 & -1 \end{bmatrix}$ | $\begin{bmatrix} 0 & 0 & 0 \end{bmatrix}$ |
| 3 | $-B^e \begin{bmatrix} 0 & 1 & -1 \end{bmatrix}$ | $B^e \begin{bmatrix} 1 & 0 & 1 \end{bmatrix}$ | $\begin{bmatrix} 0 & 0 & 0 \end{bmatrix}$ |
| 4 | $B^e \begin{bmatrix} 0 & 1 & 1 \end{bmatrix}$ | $B^e \begin{bmatrix} 1 & 0 & 1 \end{bmatrix}$ | $\begin{bmatrix} 0 & 0 & 0 \end{bmatrix}$ |
| 5 | $-B^e \begin{bmatrix} 0 & 1 & 1 \end{bmatrix}$ | $B^e \begin{bmatrix} 1 & 0 & -1 \end{bmatrix}$ | $\begin{bmatrix} 0 & 0 & 0 \end{bmatrix}$ |
| 6 | $B^e \begin{bmatrix} 0 & 1 & -1 \end{bmatrix}$ | $B^e \begin{bmatrix} 1 & 0 & -1 \end{bmatrix}$ | $\begin{bmatrix} 0 & 0 & 0 \end{bmatrix}$ |
| 7 | $B^e \begin{bmatrix} 0 & 1 & -1 \end{bmatrix}$ | $-B^e \begin{bmatrix} 1 & 0 & 1 \end{bmatrix}$ | $\begin{bmatrix} 0 & 0 & 0 \end{bmatrix}$ |
| 8 | $-B^e \begin{bmatrix} 0 & 1 & 1 \end{bmatrix}$ | $-B^e \begin{bmatrix} 1 & 0 & 1 \end{bmatrix}$ | $\begin{bmatrix} 0 & 0 & 0 \end{bmatrix}$ |

**Table S3.4**: The effective field corresponding to the $\mathcal{T}$-even torque in absence of spin-orbit coupling for circularly polarized light propagating along the $z$ direction. $\mathbf{e} = [1, i, 0]$.

When spin-orbit coupling is included, the symmetry is lowered considerably. For the effective field corresponding to the $\mathcal{T}$-odd torque we find:

$$\mathbf{B}_1 = \begin{bmatrix} \sin^2(\alpha) \operatorname{Im}(x_{011}) + \sin(2\alpha) \operatorname{Im}(x_{010}) + \cos^2(\alpha) \operatorname{Im}(x_{000}) \\ \sin^2(\alpha) \operatorname{Im}(x_{111}) + \sin(2\alpha) \operatorname{Im}(x_{110}) + \cos^2(\alpha) \operatorname{Im}(x_{100}) \\ \sin^2(\alpha) \operatorname{Im}(x_{211}) + \sin(2\alpha) \operatorname{Im}(x_{210}) - \cos^2(\alpha) \operatorname{Im}(x_{100}) \end{bmatrix}, \tag{S3.8}$$

$$\mathbf{B}_2 = \begin{bmatrix} -\sin^2(\alpha) \operatorname{Im}(x_{100}) - \sin(2\alpha) \operatorname{Im}(x_{110}) - \cos^2(\alpha) \operatorname{Im}(x_{111}) \\ -\sin^2(\alpha) \operatorname{Im}(x_{000}) - \sin(2\alpha) \operatorname{Im}(x_{010}) - \cos^2(\alpha) \operatorname{Im}(x_{011}) \\ -\sin^2(\alpha) \operatorname{Im}(x_{100}) + \sin(2\alpha) \operatorname{Im}(x_{210}) + \cos^2(\alpha) \operatorname{Im}(x_{211}) \end{bmatrix}, \tag{S3.9}$$

$$\mathbf{B}_3 = \begin{bmatrix} \sin^2(\alpha) \operatorname{Im}(x_{211}) - \frac{\sin(2\alpha) \operatorname{Im}(x_{121})}{2} + \frac{\sin(2\alpha) \operatorname{Im}(x_{221})}{2} - \cos^2(\alpha) \operatorname{Im}(x_{111}) \\ \sin^2(\alpha) \operatorname{Im}(x_{111}) + \frac{\sin(2\alpha) \operatorname{Im}(x_{121})}{2} - \frac{\sin(2\alpha) \operatorname{Im}(x_{221})}{2} - \cos^2(\alpha) \operatorname{Im}(x_{211}) \\ \sin(2\alpha) \operatorname{Im}(x_{021}) + \operatorname{Im}(x_{011}) \end{bmatrix}. \tag{S3.10}$$

Here $x_{ijk}$ denote free parameters and $\alpha$ is the in-plane polarization angle. Note that this effective field is directly determined from symmetry analysis and we do not enforce the condition that $\mathbf{B}$ is perpendicular to $\mathbf{T}$ because doing so results in a much more complicated expression. This does not matter in practice since only the perpendicular component matters when the torque is evaluated from the effective field and furthermore, we find that the direction of the effective field is not symmetry-constrained in this case.

In the following section we will show how $\mathcal{T}$-odd torque in this case could be understood as a spin-transfer torque resulting from electron excitation described by Fermi's golden rule of light absorption in non-collinear antiferromagnets.

# 4 Supplementary Note 4. Laser-induced torque due to electrons photoexcited from neighbouring sublattices

Interaction with photons from the pump laser leads typically to electron excitations from the 3$d$ band to $sp$-like bands and creates a population of excited electrons at more delocalized levels [22]. Direct consequences of this type of electron excitation on magnetization have been studied in the context of ultrafast demagnetization where the difference of lifetimes for different spins has been found to play a crucial role [23, 24]. There is an evidence for transfer of spins between different atoms in optically pumped systems [25]. Most studies of this effects were limited to collinear magnets so far. Such transfer of electrons could in noncollinear systems also generate a torque, as shown for magnetic junctions with perpendicularly oriented layers [26, 27]. The effect could be applicable also to system composed of noncollinearly oriented sublattices, however, we are not aware of any related study. Here we consider this type of torque for Mn$_3$NiN, the microscopic mechanism that could be behind it and examine whether such type of torque could be responsible for the observed magnetization dynamics.

## 4.1 Antiferromagnet Mn$_3$NiN with three magnetic sublattices

In Mn$_3$NiN three different Mn magnetic sublattices could be excited. The torque on one sublattice, e.g., *1*, is given by excited electrons from the other two sublattices *2*, *3*. Their spin orientation differ from that of *1* by the angle of $\pm 120°$. If the contributions from the other two sublattices *2* and *3* were the same, torques would cancel each other, as the effect is odd in $M$. The difference between excitation rates for the sublattices is thus the crucial quantity. We evaluate the excitation rate resolved with respect to source sublattice using first principles calculations in order to reveal their difference.

Transition rates are given by Fermi's golden rule for absorption. We assume the Hamiltonian for light-matter interaction using dipole approximation and excitation by plane wave with vector potential $\mathbf{A} = A_0 \boldsymbol{e} \exp\left[i(\mathbf{q}.\mathbf{r} - \omega t)\right]$, where $\boldsymbol{e}$ is the unit polarization vector in the $xy$ plane and $A_0$ is the amplitude of the wave. For the transition rate $w_{QQ'}$ between the eigenstates $\phi_\mathbf{k}^{Q'}$ and $\phi_\mathbf{k}^{Q}$ we obtain

$$w_{QQ'}(\boldsymbol{k}, \epsilon) = \frac{2\pi}{\hbar} \left| \left\langle \phi_\mathbf{k}^{Q'} \left| \frac{e}{m_e c} A_0 \boldsymbol{e} \cdot \boldsymbol{p} \right| \phi_\mathbf{k}^{Q} \right\rangle \right|^2 \delta\left(\epsilon_{Q'}(\boldsymbol{k}) - \epsilon_Q(\boldsymbol{k}) - \epsilon\right), \tag{S4.1}$$

where $\epsilon = \hbar\omega$ is the photon energy, $\epsilon_Q(\boldsymbol{k})$ is the energy of the $Q$-th band with crystal momentum $\boldsymbol{k}$, $e$ is the electron charge, $m_e$ is its mass, $\boldsymbol{p}$ is its momentum, $c$ is the speed of light and $Q$, $Q'$ are the band indices. Assuming zero temperature we require $\epsilon_{Q'}(\boldsymbol{k'}) > \epsilon_F$, $\epsilon_Q(\boldsymbol{k}) < \epsilon_F$, where $\epsilon_Q(\boldsymbol{k})$ and $\epsilon_{Q'}(\boldsymbol{k'})$ are the energies of the initial and final state, respectively, and $\epsilon_F$ is the Fermi energy.

It is possible to project the eigenstates on the basis of sublattices $n$ ($n = 1, 2, 3$): $\left| \phi_\mathbf{k}^{Q,n} \right\rangle = |n\rangle \langle n| \phi_\mathbf{k}^{Q} \rangle$ in order to resolve the rate of photoexcitation of electrons with respect to which sublattice $n$ the initial state belongs to. Sublattice-resolved total photoexcitation rate $w_e^n(E)$ is then given as

$$w_{\boldsymbol{e}}^n(\epsilon) \propto \sum_{Q'Qk} \frac{2\pi}{\hbar} \left| \left\langle \phi_\mathbf{k}^{Q'} \left| A_0 \boldsymbol{e} \cdot \boldsymbol{p} \right| \phi_\mathbf{k}^{Q,n} \right\rangle \right|^2 \delta\left(\epsilon_{Q'}(\boldsymbol{k}) - \epsilon_Q(\boldsymbol{k}) - \epsilon\right). \tag{S4.2}$$

The density of electrons excited by pump with polarization $\boldsymbol{e}$ at a site $n$ is proportional to $w_{\boldsymbol{e}}^n(\epsilon)$. This quantity depends significantly on parameters $n$ and $\boldsymbol{e}$, i.e., on which Mn sublattice is being excited and on the polarization of light. If we assume that cubic symmetry is present, i.e., sample is not strained, and spin-orbit coupling (SOC) is neglected, it acquires a particularly simple form which depends on the parameters $n$, $\boldsymbol{e}$ solely via the angle $\vartheta_n$ of the polarization w.r.t. sublattice-specific vectors $\boldsymbol{\varepsilon}_n$ (see Fig. S17). Regardless of which of the eight possible $\Gamma^{4g}$ magnetic domains we consider (see Fig. S18), these sublattice-specific vectors align along local uniaxial symmetry axes: $\boldsymbol{\varepsilon}_1 = \boldsymbol{x}$, $\boldsymbol{\varepsilon}_2 = \boldsymbol{y}$, $\boldsymbol{\varepsilon}_3 = \boldsymbol{z}$. This allows us to define $\nu(\epsilon) = w_{\boldsymbol{\varepsilon}_n}^n(\epsilon)$, $\delta\nu(\epsilon) = w_{\boldsymbol{l}}^n(\epsilon) - w_{\boldsymbol{\varepsilon}_n}^n(\epsilon)$ and then the excitation rate can be expressed as:

$$w_{\boldsymbol{e}}^n(\epsilon) = \nu(\epsilon) + \delta\nu(\epsilon) \sin^2 \vartheta_n, \tag{S4.3}$$

where $\boldsymbol{l}$ is a unit vector perpendicular to both chosen sublattice-specific vector $\boldsymbol{\varepsilon}_n$ and the direction of light propagation $\mathbf{q}$.

For the geometry that we consider here, i.e., light is incident along the $\boldsymbol{z}$ direction and polarized in the $xy$ plane (see Fig. S17 and S18), the following relations hold: $\boldsymbol{e} = (\cos\alpha, \sin\alpha, 0)$, $\vartheta_1 = \alpha$, $\vartheta_2 = \alpha - 90°$, $\vartheta_3 = 90°$. And as shown in Fig. S9b, the PPD signal is pretty much independent of $AOI_{pump}$ within the range of $\pm 15°$, therefore, the normal incidence of the pump light within this model is justified.

In a more realistic scenario, where SOC is finite and sample is strained, the relation (S4.3) still holds true. However, the directions of local uniaxial symmetry axes are not along the $\boldsymbol{x}$, $\boldsymbol{y}$ and $\boldsymbol{z}$ coordinates and the sublattice-specific vectors $\varepsilon_n$ are tilted. For example, in the case of Domain 2 with no strain, $\varepsilon_1 = (0.99, 0.08, 0.08)$, $\varepsilon_2 = (-0.08, 0.99, -0.08)$, $\varepsilon_3 = (0.08, 0.08, 0.99)$. Though, as we stated in the main text and as we simulate dynamics in Sec. 4.5, taking into account SOC changes the outcome of dynamical simulations (and therefore calculated MO response) only insignificantly.

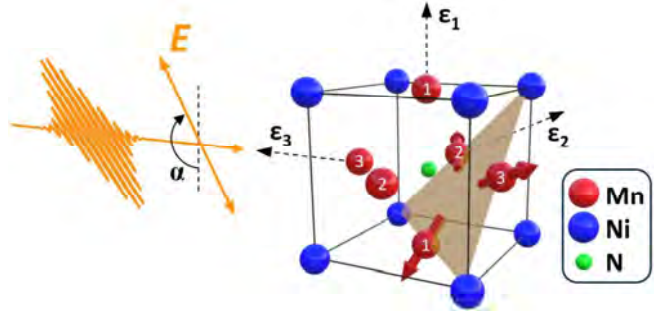

**Fig. S17**: **Geometry for torque due to excited electrons from neighbouring sublattices.** Schematic illustration of the geometry for inducing magnetic dynamics in Mn$_3$NiN via torques from neighbouring sublattices. Pump pulses are incident normal to the atomic plane. $\varepsilon_n$ are the sublattice-specific vectors. For simplicity, only three of the six spins from the unit cell are shown. Domain 1.

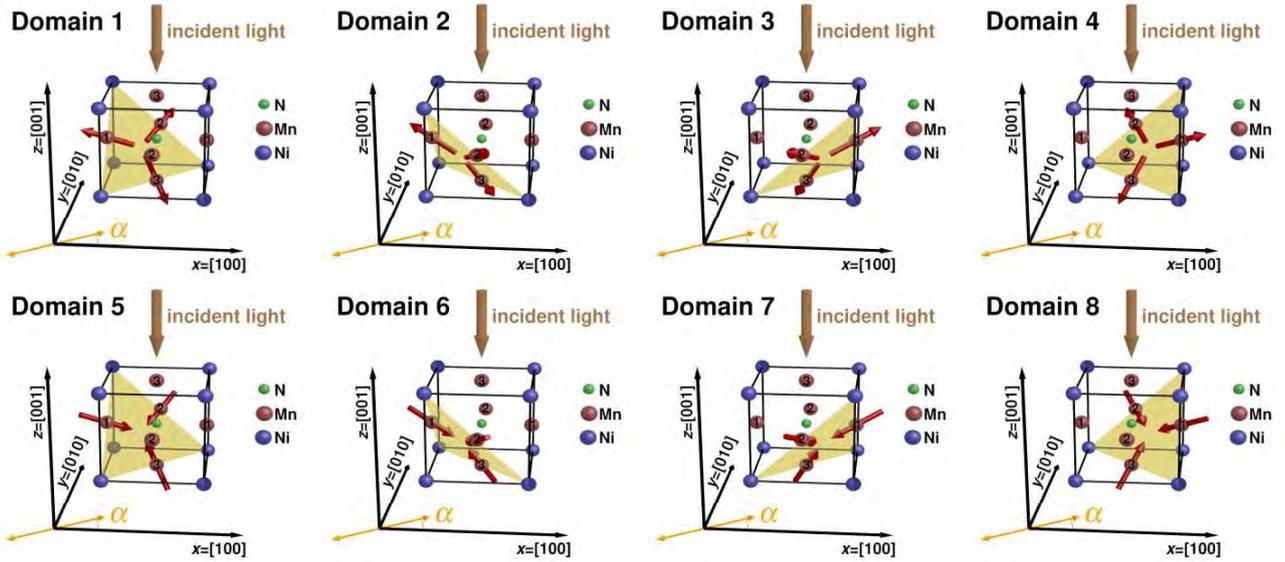

**Fig. S18**: $\Gamma^{4g}$ **Domains.**, The eight variants of $\Gamma^{4g}$ magnetic structure and the direction of light propagation. Incident light, pump, is along the $z$ axis and is polarized in the $xy$ plane at an angle $\alpha$. For simplicity, only three of the six spins from the unit cell are shown.

For a chosen pump photon energy $\epsilon$, the quantities $\nu(\epsilon)$ and $\delta\nu(\epsilon)$ are constants and $w_e^n(\epsilon)$ depends only on $\alpha$, the orientation of linear polarization of the pump light. On the other hand, for a constant $\alpha$, the rates $\nu$ and $\delta\nu$ depend on the energy of the exciting photons. These dependencies are shown in Fig. S19 for the cases with and without SOC. However, here we must stress that these calculations of $\nu(\epsilon)$ and $\delta\nu(\epsilon)$ do not take into accounts all the possible contributions to the electronic excitation and they are, as every theoretical description, approximate. In our experiments we saw virtually no dependence of magnetic dynamics on the pump wavelength in the range of $\lambda \in [800\,\text{nm}, 870\,\text{nm}]$, which corresponds to photon energies approximately $\epsilon \in [1.42\,\text{eV}, 1.55\,\text{eV}]$. Fig. S19 shows the energy dependence of relative strengths of the rates $\nu$ and $\delta\nu$. As expected for metallic compounds, the main contribution comes from the polarization-independent part, $\nu(\epsilon)$. On the other hand, after summing the contributions from all three sublattices to the total excited spin angular momentum, this part is suppressed, while the polarization-dependent part, $\delta\nu(\epsilon)\sin^2\vartheta_n$, ends up non-zero.

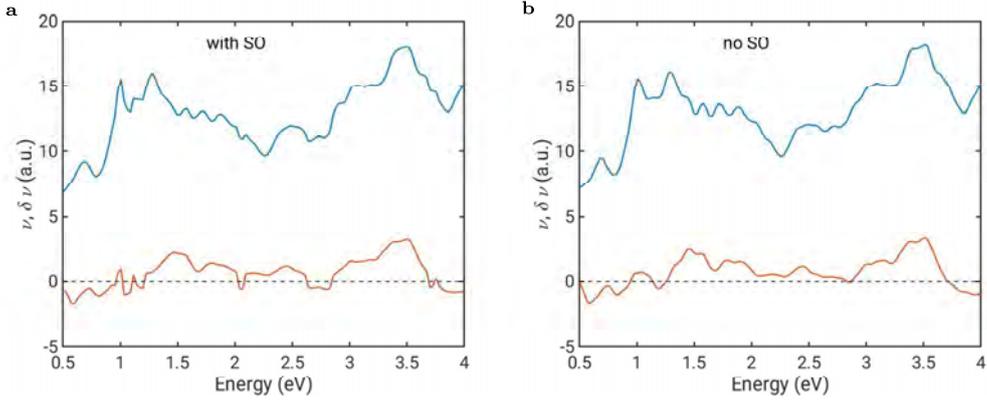

**Fig. S19**: **Density of excited electrons.** Calculated density of excited electrons at sublattice 1 in Mn$_3$NiN as a function of pump photon energy. Blue curve: $\nu$, red curve: $\delta\nu$. **a**, With SOC. **b**, Without SOC.

## 4.2 Bandstructure and polarization dependence of sublattice excitation rates

In order to better understand the polarization dependence of sublattice excitation rates, we plot the bandstructure in Fig. S20 with an indication of the contributions of individual Mn sublattices. Notably the bandstructure of Mn$_3$NiN is invariant w.r.t. the $C_3$ axis rotations, but the contributions of individual Mn sublattices are not. The three Mn sublattices also exhibit significantly different occupations of $d$ orbitals due to the different orientation of their neighbours (ligands), therefore one could expect different response of Mn sublattices to the perturbation by light with different polarization.

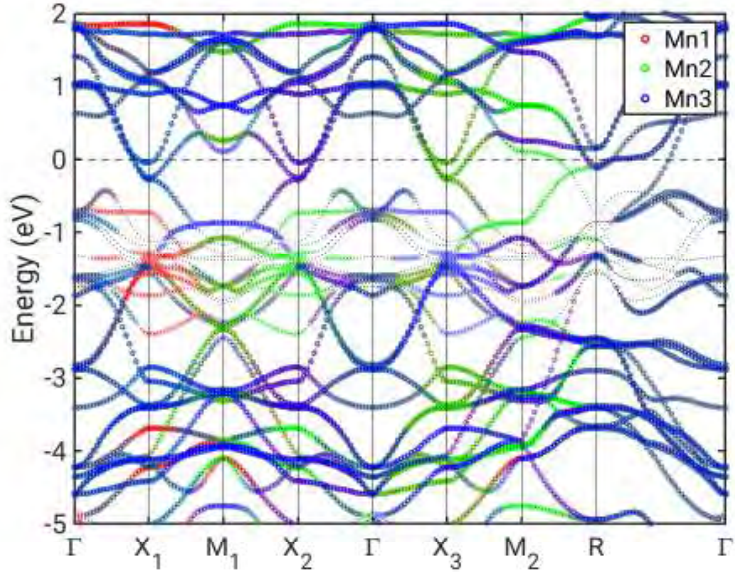

**Fig. S20**: Mn$_3$NiN bandstructure with proportions of individual Mn sublattice contributions shown by selected colors. BZ points X1, X2, X3 correspond to X points located along the x, y and z axis, respectively.

## 4.3 Torques

The torque $\boldsymbol{\tau}_n$ exerted on local magnetic moment $\boldsymbol{s}_n$ (with unit length) at sublattice $n$ in an antiferromagnet with three sublattices is given by contributions of excited electron from the other 2 sublattices. Since the number of electrons excited at site $n'$ is proportional to $w_e^{n'}$ and their spins are along $\boldsymbol{s}_{n'}$, the torque could be written as [28]

$$\boldsymbol{\tau}_n = A_{\text{ncol}} \boldsymbol{s}_n \times \left[ \boldsymbol{s}_n \times \left( w_e^{n'} \boldsymbol{s}_{n'} + w_e^{n''} \boldsymbol{s}_{n''} \right) \right]. \tag{S4.4}$$

The proportionality constant $A_{\text{ncol}}$ contains multiple effects. It is not possible to calculate it within our theory, but its knowledge is not needed because we are not interested in total magnitude, just in dependences on polarization angles.

Unlike the sublattice-specific vectors, which do not depend on the magnetic domain (when SOC is zero), the spin vectors $s_n$ naturally do (see Table S4.1), therefore the torque $\tau_n$ does, too. Despite this, its direction is in the plane of the spins no matter which domain, therefore one could expect that this torque tends to tilt the local magnetic moments closer to or away from each other in the beginning of the dynamics. The case of Domain 1 without SOC is illustrated in Fig. S21 and in Fig. 4a in the main text. For finite SOC, when the sublattice-specific vectors are tilted, the excitation rates are changed slightly and therefore torques change, too. We illustrate this in Fig. S22 and in Fig. 4d in the main text. In particular, one can see that the torques for light polarization directions $\alpha = 45$° and $135$° are no longer degenerate.

Table S4.1: Directions of local magnetic moments $s_n$ for eight variants of $\Gamma^{4g}$. Non-normalized.

| Domain # | $s_1$ | $s_2$ | $s_3$ |
| --- | --- | --- | --- |
| 1 | (-2, -1, 1) | (1, 2, 1) | (1, -1, -2) |
| 2 | (-2, 1, 1) | (1, -2, 1) | (1, 1, -2) |
| 3 | (2, 1, 1) | (-1, -2, 1) | (-1, 1, -2) |
| 4 | (2, -1, 1) | (-1, 2, 1) | (-1, -1, -2) |
| 5 | (2, 1, -1) | (-1, -2, -1) | (-1, 1, 2) |
| 6 | (2, -1, -1) | (-1, 2, -1) | (-1, -1, 2) |
| 7 | (-2, -1, -1) | (1, 2, -1) | (1, -1, 2) |
| 8 | (-2, 1, -1) | (1, -2, -1) | (1, 1, 2) |

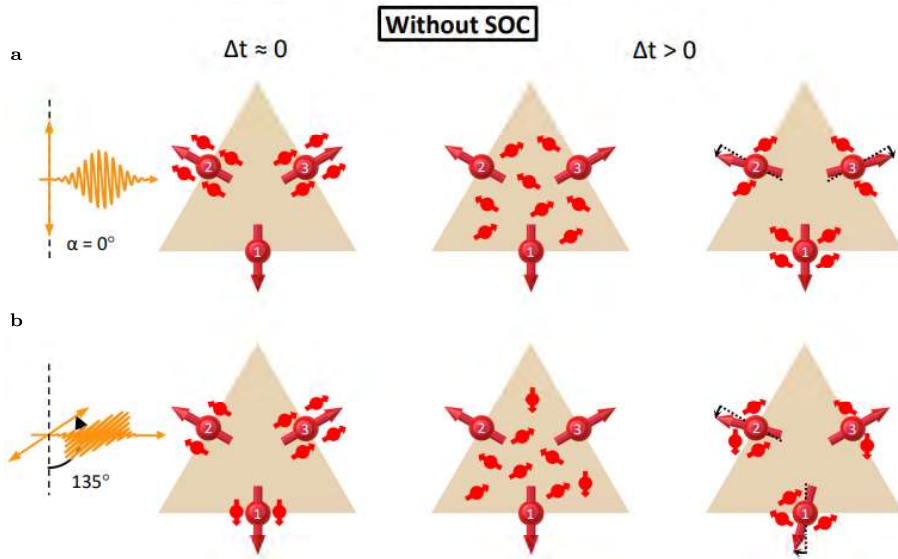

Fig. S21: **Torque due to excited electrons from neighbouring sublattices without SOC. a,** Schematics of the microscopic mechanism for inducing magnetic dynamics with pump linear polarization oriented along $\alpha = 0$°. Pump pulse excites various numbers of electrons at Mn sites at $\Delta t \approx 0$. Density of excited electrons from $n$-th Mn site depends on the angle between pump electric field and $\varepsilon_n$. Due to finite conductivity of the material, electrons move around freely and interact with localized electrons at other Mn sites. This results in torque. **b,** As in **a**, but $\alpha = 135$°. Domain 1.

To summarize the mechanism: the excitation rates are generally not equivalent for all 3 sublattices. Spin contributions to the excited state populations then do not cancel each other, this population is in total spin polarized, and capable of exerting torque on localized electrons. Furthermore, the difference of the number of excited electrons for different sublattices depends on the pump polarization and hence the torque depends on the pump polarization. We have shown that in a noncollinear AFM system one can achieve a small non-zero total spin polarization of excited electrons and tune its orientation by the orientation of pump polarization. The orientation of pump polarization therefore allows to control which Mn sublattice is predominantly excited.

**Torque as a result of spin-polarized current**

The expression (S4.4), from [28], was derived for a spin-transfer torque exerted in heterostructures in the limit when the spin-mixing conductance is dominated by its real part. Assuming that spin is conserved when the electrons are excited, we could understand the spin-polarized population of excited electrons as a spin-polarized current flowing to one site from other sites. The spin current $\boldsymbol{p}_n$ flowing to site $n$ will be polarized along the direction

$$\boldsymbol{p}_n \propto \boldsymbol{s}_{n'} \sin^2 \vartheta_{n'} + \boldsymbol{s}_{n''} \sin^2 \vartheta_{n''} \tag{S4.5}$$

and could be seen as a spin-transfer torque (STT).

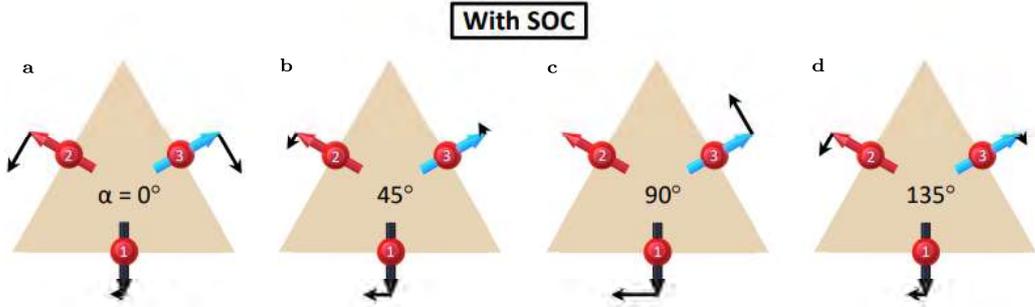

**Fig. S22**: **Torque due to excited electrons from neighbouring sublattices with SOC.** Torques induced on three spin sublattices (black arrows) by pump pulses with various linear polarizations. Non-zero SOC. Red circles represent Mn atoms and corresponding arrows their spins. Domain 1. **a**, $\alpha = 0°$. **b**, $\alpha = 45°$. **c**, $\alpha = 90°$. **d**, $\alpha = 135°$.

In a ferromagnet with magnetization $\boldsymbol{M}$, STT induced by a spin current with spin polarization $\boldsymbol{p}$ acting on $\boldsymbol{M}$ has two component: field-like (FL) and anti-damping-like (AD):

$$\boldsymbol{\tau}^{FL} = \boldsymbol{M} \times \boldsymbol{p}, \tag{S4.6}$$

$$\boldsymbol{\tau}^{AD} = \boldsymbol{M} \times (\boldsymbol{M} \times \boldsymbol{p}) \tag{S4.7}$$

with corresponding effective fields

$$\boldsymbol{B}_{STT}^{FL} = \boldsymbol{p}, \tag{S4.8}$$

$$\boldsymbol{B}_{STT}^{AD} = \boldsymbol{M} \times \boldsymbol{p}. \tag{S4.9}$$

Assuming that this will work similarly in the antiferromagnet for each sublattice $n$ with local magnetic moment $\boldsymbol{s}_n$, we again get two torque components described by effective fields:

$$\boldsymbol{B}_{STT,n}^{FL} = \boldsymbol{p}_n, \tag{S4.10}$$

$$\boldsymbol{B}_{STT,n}^{AD} = \boldsymbol{s}_n \times \boldsymbol{p}_n. \tag{S4.11}$$

Equations (S4.7), (S4.9) and (S4.11) for the AD torque and corresponding effective fields have the same symmetry and directions as the $\mathcal{T}$-odd torque and corresponding effective fields in Sec. 3 (Eq. (S3.6) and Table S3.1). For this reason, we will be referring to the $\mathcal{T}$-odd torque as AD torque, too. The same connection is between the $\mathcal{T}$-even torque and FL torque.

## 4.4 Permittivity tensor

To evaluate the magneto-optical (MO) response of our system, i.e., the change of ellipticity or rotation of probe light transmitted through the sample at a given time after excitation by pump pulse, we need to calculate the permittivity tensors for both the ground state (before pump) and the excited state (after pump). We chose to evaluate the MO response of the system $\Delta t = 3$ ps after the excitation. This is safely after $\Delta t = 1.5$ ps, the time delay until which the dynamics of rotation and ellipticity in Mn$_3$NiN differ (see Fig. 2b, Fig. S12), likely due to transient optical effects.

In order to calculate the permittivity tensor as a function of energy, we first use non-collinear spin-polarized density functional theory (DFT) to find the electronic structure. The spin-orbit coupling is included and the local magnetic moments are constrained to either $\Gamma^{4g}$ configuration (ground state) or directions predicted by the dynamical model described in the next section. Subsequently, we employ linear response theory following the approach of [29] and our previous work [30, 31]. We use the projector augmented wave method implemented in the VASP code [32] with generalized gradient approximation (GGA) parametrized by Perdew–Burke–Ernzerhof [33]. The plane-wave energy cutoff is 500 eV. The valence configurations of manganese, nickel and nitrogen are $3p^6 3d^6 4s^1$, $3p^6 3d^9 4s^1$, and $2s^2 2p^3$, respectively. We use a regular $20 \times 20 \times 20$ reciprocal space mesh.

To simulate properties of the measured films more accurately, we include a biaxial lattice strain induced during growth due to a lattice mismatch with the substrate. Our lattice parameters are: $a = b = 0.3839$ nm and $c = 0.3846$ nm (that corresponds to a compressive strain, $\varepsilon_{xx} = -0.06\%$, with respect to the relaxed lattice parameter $a = 0.38412$ nm). The penalty term is set to $\lambda = 10$ which is sufficiently high to suppress the canting of the moments in strained unit cell due to the piezomagnetic effect [1]. Further details related to calculating the permittivity tensor for Mn$_3$NiN could be found in [30].

## 4.5 Magnetic dynamics

We simulate the magnetic dynamics using the Landau-Lifschitz-Gilbert equations [34]. For simplicity, we assume a macrospin model, which corresponds to a single domain with homogeneous magnetic dynamics. We use a version from [35], which consists of isotropic Heisenberg exchange and uniaxial anisotropy. The model was derived for Mn$_3$Ir, which has the same symmetry as Mn$_3$NiN. It can be written as:

$$H = J(\mathbf{S}_1 \cdot \mathbf{S}_2 + \mathbf{S}_2 \cdot \mathbf{S}_3 + \mathbf{S}_1 \cdot \mathbf{S}_3) - K/2(S_{1,x}^2 + S_{2,y}^2 + S_{3,z}^2) \,. \tag{S4.12}$$

To estimate the parameters of the model, we use VASP calculations as described in the previous section. By tilting the magnetic moments with respect to each other, we get $J = 291$ meV. We evaluate the magnetic anisotropy by rotating the magnetic moments in the plane. From this we determine that $K = 0.1$ meV. Note that the main results from the dynamics simulations do not depend on the exact values of these parameters and that we are considering here the unstrained case. The key aspect is that the exchange is very strong and much stronger than the anisotropy.

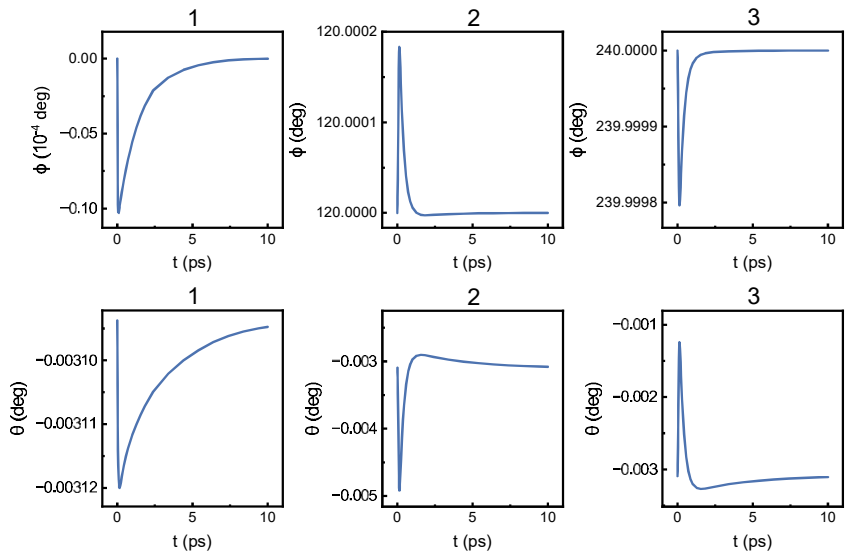

**Fig. S23**: **Simulations for the field-like torque.** Results of the dynamics simulations for the FL torque, polarization angle $\alpha = 0°$ and effective field magnitude 10 mT. 1, 2, 3 correspond to different sublattices and angles $\phi$ and $\theta$ denote the orientation with respect to the equilibrium plane of the magnetic moments as described in the text.

To solve the LLG equations we use the torques obtained from the microscopic calculations (Sec. 4.3). In the absence of spin-orbit coupling, essentially the same result would be obtained using the torque determined from symmetry analysis (Sec. 3) since it contains no parameters apart from the overall magnitude and we only consider small deviations from equilibrium. We set the Gilbert damping parameter to 0.1. This does not have a strong influence on the induced magnetic dynamics in our case. To simulate the effect of the optical pulse we consider a time-dependent torque with a very fast rise and a relaxation on a ps timescale. We model the pulse by a linear rise to maximum torque amplitude at $t = 100$ fs and then exponential relaxation with relaxation time 0.3 ps. This is chosen so that the relaxation of the spin dynamics roughly correspond to the experimentally observed relaxation.

The results of the simulations for Domain 1 and the FL torque are given in Fig. S23. Here the magnitude of the torque is chosen such that the effective field for Domain 1, sublattice 2 and $\alpha = 0°$ is 10 mT. The plot uses a coordinate system defined with respect to the plane in which the magnetic moments lie in equilibrium. The coordinate system is chosen such that $\phi$ measures the angle from the equilibrium moment 1 direction within the plane and $\theta$ measures the tilting out of this plane. These results show that the FL torque has only a very small effect on the magnetic state. This is the expected result since as shown in Table S3.2, the effective field drives the magnetic moments against the exchange interaction, which is very strong. An analogous plot for the AD torque is given in Fig. S24. We see that here the deviation from the equilibrium state is much larger even though the torque magnitude is the same. This is because in this case, the effective field acts against the anisotropy rather than the exchange.

In Fig. S25 we show the result for the AD torque for 45° and 135° polarization directions with and without SOC. As expected we find that without SOC the results are exactly the same. With SOC, they differ somewhat though the difference is quite small.

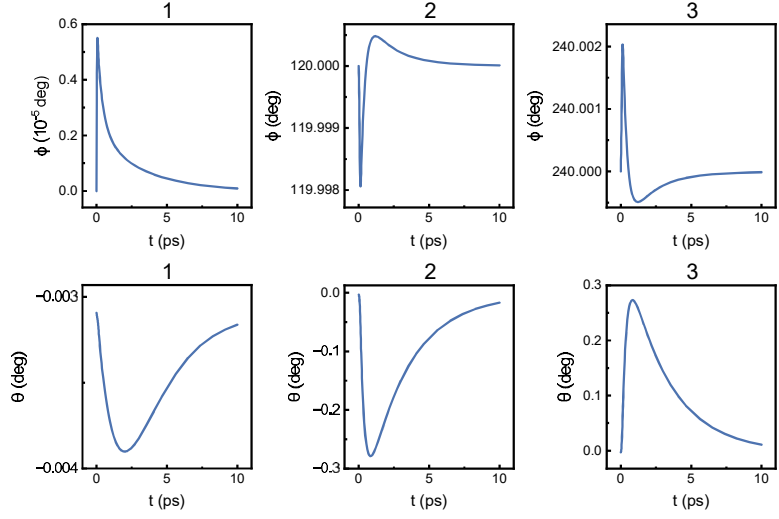

**Fig. S24**: **Simulations for the anti-damping-like torque.** As in Fig. S23, but for AD torque.

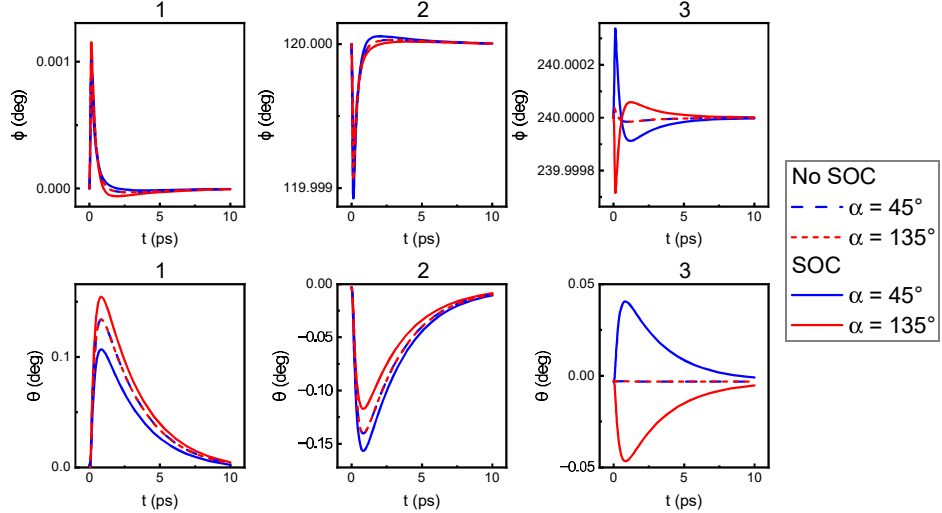

**Fig. S25**: **AD torque, SOC coupling.** Comparing the results of the dynamics simulations for the AD torque with and without spin-orbit coupling and for polarization angles $\alpha = 45°$ and $\alpha = 135°$. The effective field magnitude is 10 mT.

### 4.6 Magneto-optical response

Using the Yeh's formalism [36] with the permittivity tensors for the ground state and the excited states, we calculated the MO response of the 13 nm thin film of $Mn_3NiN$. Within these calculations, we analysed the MO response of all possible 8 domains. We found out that Domains 1 and 3 generate the same MO signals. The same is true for domain pairs 2 and 4, 5 and 7, and 6 and 8. This is because the domains from a pair are related with rotation by 180° around the [001] direction (see Fig. S18). For this reason, we only need to take into account Domains 1, 2, 5 and 6. After averaging over these domains, we find the dependence on the probe polarization to be in very good agreement with the experimental observations for the pump polarization $\alpha = 0°$ (see Fig. S7b and Fig. 4c in the main text).

This model based on the Fermi's golden rule, however fails to reproduce the full dependence on the pump polarization. It predicts the same excitation rates for pump polarizations along the directions $\alpha = 45°$ and $\alpha = 135°$, when the SOC is neglected (see Eq. (S4.3)). Finite SOC lifts this degeneracy (see Fig. S22 for illustration), however it still does not predict opposite MO response, which we observe in the experiment (see Fig. 1b).

# 5 Supplementary Note 5. Real time quantum-classical dynamics of non-collinear magnet and formation of spin spiral

Because the experimentally observed dynamics of pump-induced changes of magnetic-ordering cannot be sufficiently explained by the second-order torque considering a primitive cell, as described in the previous sections, we now turn to a different type of excitation mechanism taking into account multiple primitive cells.

The real time dynamics of the non-collinear spin system is studied with a hybrid quantum-classical time evolution formalism [37, 38]. This formalism has been quite successful in exploring optically induced magnetic properties [39, 40]. In this framework, the dynamics of the electronic subsystem is treated quantum mechanically, whereas the dynamics of the magnetic subsystem is treated classically. The magnetic interaction among the localised magnetic moments is mediated by the itinerant electrons which is defined by a double-exchange model given by

$$H = -J_M \sum_n a_n^\dagger (\sigma \cdot \mathbf{s}_n) a_n - h_t \sum_{\langle n,n' \rangle} a_n^\dagger a_{n'} \ , \tag{S5.1}$$

where $J_M$ is the Hund coupling between the localised magnetic moment and the itinerant electrons. $\mathbf{s}_n$ is the unit vector denoting the direction of the localized magnetic moment at site $n$ (see Fig. S26a,b) and $\sigma$ is the vector of Pauli matrices. $h_t$ is the hopping parameter. For our study we consider a $10 \times 10$ supercell of a Kagome lattice with $J_M = 1$ and $h_t = 0.4$. For simplicity, here we consider only the nearest neighbour hopping ($\langle n, n' \rangle$). $a_n^\dagger / a_n$ represents the electronic creation/annihilation operator at site $n$. The laser pulse is treated as a classical electric field given by

$$\mathcal{E}(t) = \mathcal{E}_0 \mathbf{e} \exp\left[-(t-t_0)^2/2s^2\right] \cos\left[\omega(t-t_0)\right] \tag{S5.2}$$

where $\mathcal{E}_0$ is the peak amplitude of the electric field occurring at time $t_0$. $\omega$ is the frequency of the laser and $s$ is the Gaussian width. $\mathbf{e}$ denotes the plane of polarization which can be represented by a polarization angle $\alpha$ such that $\mathbf{e} = \cos(\alpha)\mathbf{x} + \sin(\alpha)\mathbf{y}$ (see Fig. S26c,d). The laser field is incorporated within the Hamiltonian as a Peierls phase by replacing the hopping parameter $h_t$ with $h_t \exp\left[i\frac{q}{\hbar}\mathbf{A}(t) \cdot \mathbf{r}_{nn'}\right]$. Here $q$ is the electric charge, $\hbar$ is the reduced Plank's constant and $\mathbf{A}$ is the vector potential of the laser field such that $\mathcal{E} = -\partial \mathbf{A}/\partial t$. Since there is no external magnetic field, we choose $\mathbf{A}$ to be independent of spatial coordinates. $\mathbf{r}_{nn'}$ is the vector connecting site $n$ with site $n'$. For our study, we choose an electric field with peak amplitude $0.06\,\mathrm{eV}/a_0$, where $a_0$ is the distance between two nearest magnetic sites. For the Mn atoms in Mn$_3$NiN, $a_0 = 2.74\,\text{Å}$ from which we get $\mathcal{E}_0 = 2.2\,\mathrm{MV.cm}^{-1}$. We choose $s = 10\,\mathrm{fs}$ and $\omega = 2.28 \times 10^{15}\,\mathrm{Hz}$. For initial magnetic state we consider both $\Gamma^{4g}$ and $\Gamma^{5g}$ configurations with a $0.1\,\mathrm{rad}$ random fluctuation of the out of plane angle $\theta$. Assuming the pure in-plane structure of the magnetic configuration, we consider the in-plane alignment to be of $\theta = 0$ and therefore the minimum and maximum values of $\theta$ are given by $\pm 90°$.

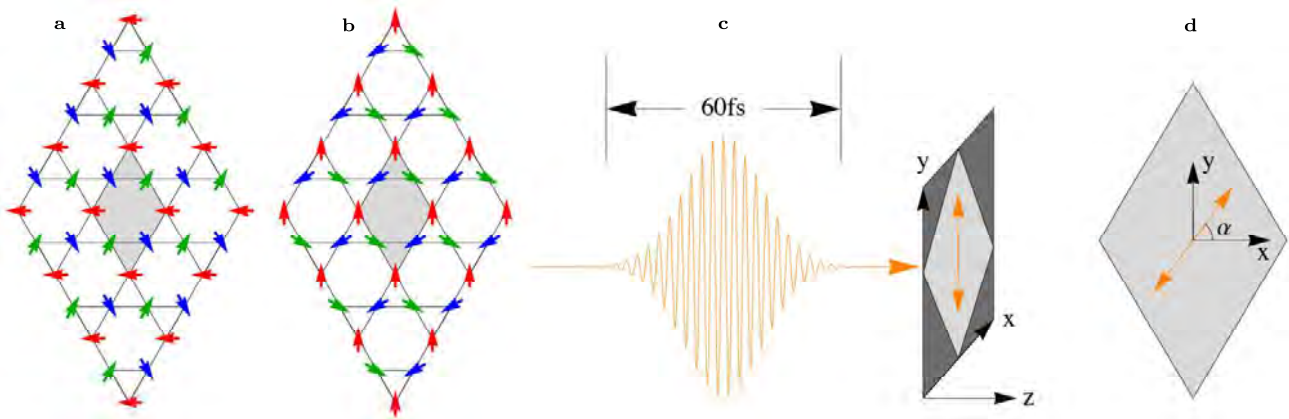

Fig. S26: **Schematic of different magnetic configuration on a kagome lattice and the laser pulse.** **a** and **b** show $\Gamma^{5g}$ and $\Gamma^{4g}$ configurations on a $3 \times 3$ super cell where red, blue and green arrows show three different sub-lattices and the gray region show the unit cell. **c** shows the laser profile for a Gaussian time width $s = 10\,\mathrm{fs}$ and **d** shows the orientation of the polarization angle $\alpha$.

The ground state of the system is constructed by filling up all the states below the Fermi level. For our choice of parameters, the valence band maxima is at $-0.2\,\mathrm{eV}$ and therefore we set our Fermi level at $-0.35\,\mathrm{eV}$ to ensure the metallic nature. The time evolution of these quantum states are described within the Schrödinger picture [37, 39]. First, we expand the states as a linear combination of the instantaneous eigenfunctions such that $\psi_l(t) = \sum c_l^m |m_t\rangle$ where $\{|m_t\rangle\}$ constitutes the complete basis of Hamiltonian $H(t)$ at time $t$. Then we use time-dependent perturbation theory to obtain the states at subsequent times. This allows us to evaluate the effective field as $\mathbf{B}_n(t) = \frac{1}{\mu_B}\langle -\partial H(t)/\partial \mathbf{s}_n \rangle_t$, where $\mu_B$ is the Bohr magneton and $\langle \rangle_t$ denotes the expectation value with respect to the states at time $t$. These effective fields are used in a set of classical Landau-Lifsitz-Gilbert equations given by

$$\frac{d\mathbf{s}_n}{dt} = -\gamma \mathbf{s}_n \times \mathbf{B}_n - \lambda \mathbf{s}_n \times (\mathbf{s}_n \times \mathbf{B}_n), \tag{S5.3}$$

where $\gamma = \frac{g_e \mu_B}{\hbar \mu_s} \frac{1}{1+\eta^2}$ and $\lambda = \frac{g_e \mu_B}{\hbar \mu_s} \frac{\eta}{1+\eta^2}$. Here $g_e$ is gyromagnetic ratio of the electron and $\mu_s$ is the saturation magnetization which we kept at $1\,\mu_B$. $\eta$ is a dimensionless constant which controls the energy dissipation of the magnetic subsystem. Here we choose $\eta = 0.05$ to ensure faster convergence. Variation of the magnitude of $\eta$ does not alter the qualitative behaviour of the final outcome [37].

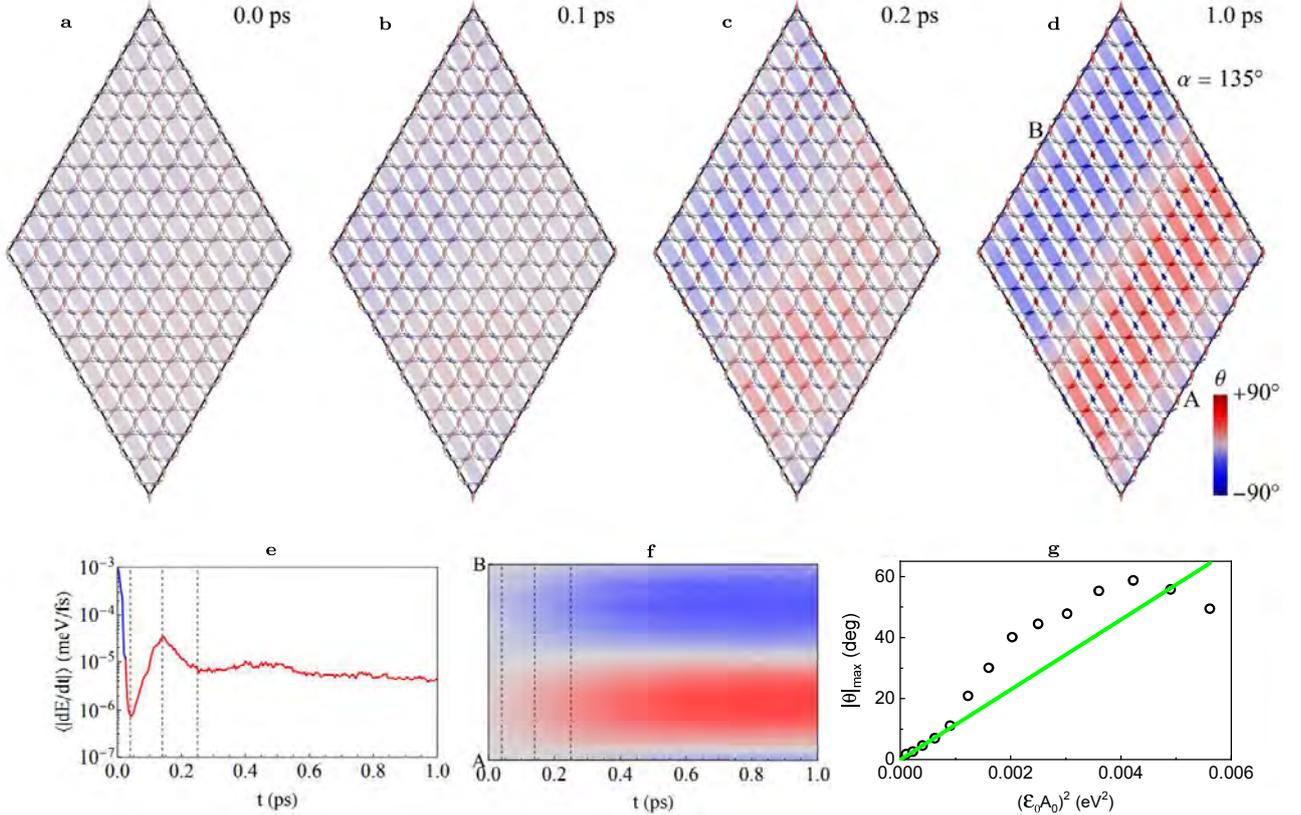

Fig. S27: **Different stages of the formation of spin spiral.** The plane of polarization of the laser ($\alpha$ in Fig. S26d) is at 135° and the peak amplitude occurs at $t = 0$. **a-d** show the spatial profile of the magnetic moments at different time, where the colour denotes the out of plane angle $\theta$ such that $\theta = 0$ represents the in-plane orientation. **e** shows the average magnitude of the energy variation where red and blue colour represent dissipation and absorption, respectively. **f** shows the time evolution of $\theta$ along a particular line marked by A-B in **d**. The vertical dashed lines at 30, 130 and 250 fs in **e** and **f** show the edge of the laser pulse and the beginning and end of the fast magnetization dynamics, respectively [37, 38]. **g** shows the variation of maximum value of $\theta$ along AB with respect to the light intensity (points), line depicts the linear dependence. See also Supplementary Video 1.

First, let us focus on a particular configuration to understand the different phases of the dynamics. From Fig. S27 and Supplementary Video 1 one can clearly see how a linearly polarized laser pulse can generate a periodic modulation parallel to the plane of polarization due to the emergent chiral interactions [37]. After excitation by the laser pulse, the system first enters a phase with a fast magnetization dynamics and then follows a slow magnetization dynamics (see Fig. S27e,f), which eventually drives it towards its metastable configuration. The energy dissipation is also decaying with time towards the convergence of the dynamics. Note that the maximum value of the out of plane angle follows a highly non-linear dependence with respect to the laser field intensity, which corresponds to the square of the light field amplitude and which is depicted as a solid line in Fig. S27g. For low intensities, the angle increases linearly, which is in accord with the experimental observations both for the $Mn_3NiN$ and $Mn_3GaN$ samples, see Fig. S10d and inset in Fig. S16, respectively. For stronger fields, the spiral formation breaks down and the system starts generating more complex magnetic textures [38], which is marked by the non-monotonic behaviour of the magnitude of $\theta$. The maximum field strength that allows the spiral formation depends on the system parameters, such as the hopping parameter, exchange parameter and the system size. The formation of the spiral, on the other hand, is a general phenomenon and can be observed for any initial magnetic configuration and for any polarization angle. To demonstrate this, we consider both $\Gamma^{5g}$ and $\Gamma^{4g}$ configuration and different polarization angles (see Fig. S28 and Supplementary Video 2).

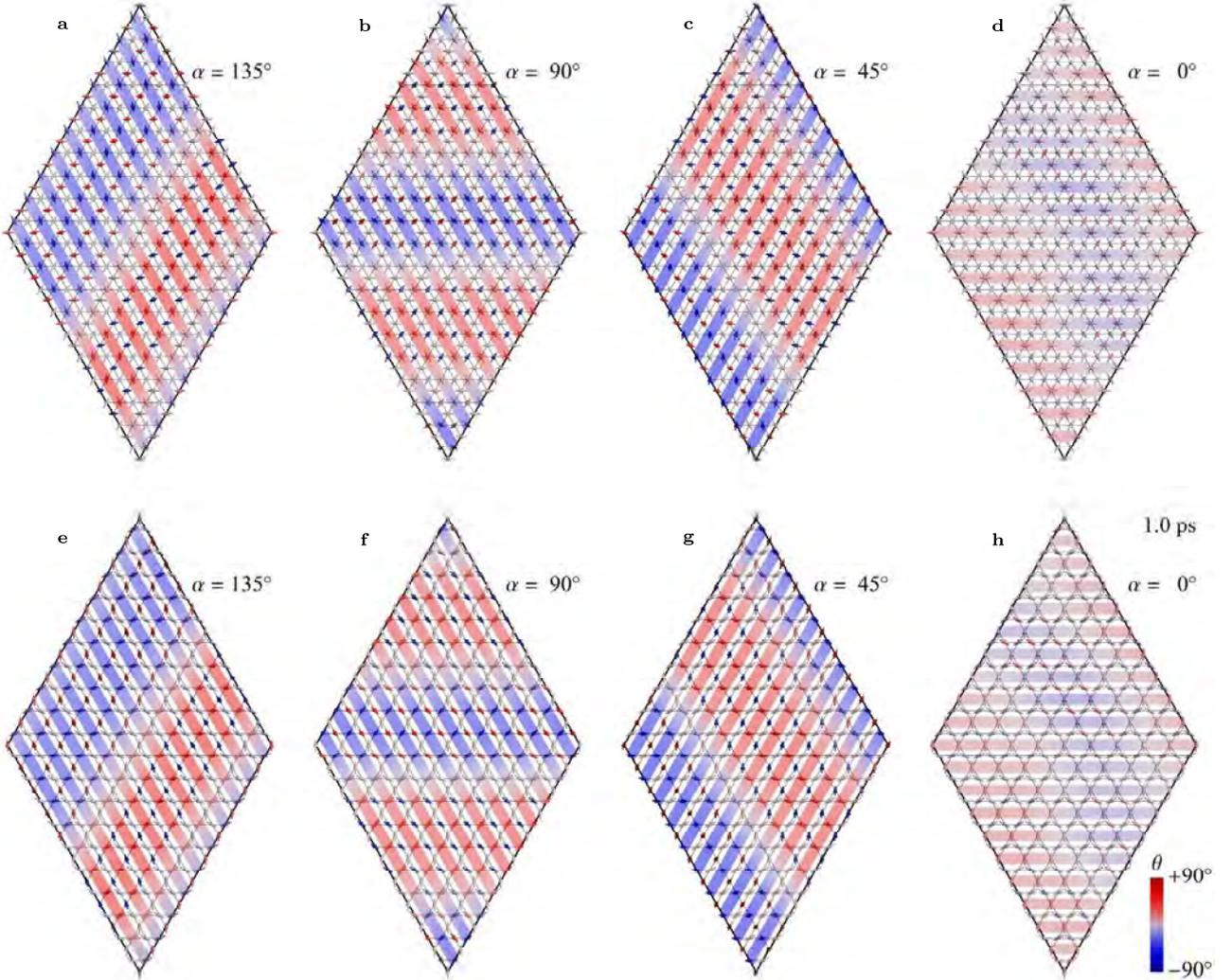

**Fig. S28**: **Magnetic configuration at 1 ps for different initial configurations and different polarizations of the laser.** For **a-d** the initial configuration is $\Gamma^{5g}$ and for **e-h** the initial configuration is $\Gamma^{4g}$. The polarization angle is set at 135° in **a** and **e**, 90° in **b** and **f**, 45° in **c** and **g** and 0° in **d** and **h**. The colour code shows the variation of the variation of the OOP angle $\theta$. See also Supplementary Video 2.

From Fig. S28 and Supplementary Video 2 one can readily see that the spiral formation always follows the plane of polarization and it is independent of the initial magnetic configuration. Note that the spiral is not very prominent when the polarization angle is 0°. This very likely originates from the small system size in our model. Due to the choice of the supercell, spin spiral along some specific directions are more stable due to the boundary effect. For large systems, such boundary effect becomes negligible and in that case spirals along any direction will be equally stable.

To address the effect of boundaries, we tested the formation of spin spiral in a system rotated in such a way that one boundary is parallel to the direction $\alpha = 0°$. Fig. 5b,c in the main text and Fig. S29 show the final magnetic configuration at 1 ps for different orientations of polarization for $\Gamma^{4g}$ configuration. The same results are obtained for $\Gamma^{5g}$ configuration (see Fig. S30). Seemingly, polarization orientations $\alpha = 0°$ and 135° produce very similar spirals. However, comparing it with Fig. S28 shows that it is an effect of a boundary and that the spirals indeed follow the plane of laser polarization.

Note that for our simulations we use a Gaussian pulse with a Gaussian time width $s = 10$ fs, which corresponds to 23.5 fs full width at half maximum (FWHM), whereas the FWHM of laser pulses duration used in the experiments is $\approx 150$ fs. Due to computational limitations, we have to use comparatively small system size, for which we also scaled the pulse width accordingly to avoid any finite size effect. To gain insight into how important this parameter is, we simulated the dynamics using the same conditions as in Fig. S27 and S28, but with various pulse durations, specifically $s = 8$ fs (see Fig. S31a,d), $s = 10$ fs (Fig. S31b,e) and $s = 13$ fs (Fig. S31c,f). Based on this, one can clearly see that the final outcome is independent of the pulse duration and therefore with a large enough system size, one could expect similar behaviour with a pulse with FWHM 150 fs as well. Interestingly, at least within the simulated pulse durations, the longer the pulse is, the sooner the spin spiral forms. This is happening because we defined the time $t = 0$ ps in the center of the pulse. Hence, at this time, the dynamics is triggered earlier by the laser rising edge for longer pulses.

The dynamics of the studied systems excited by this mechanism, as well as the symmetry of the mechanism with respect to pump polarization orientation agrees well with our experimental observations. Moreover, the fact that the predictions of this model are the same for both magnetic configurations $\Gamma^{4g}$ and $\Gamma^{5g}$ increases the credibility of this mechanism.

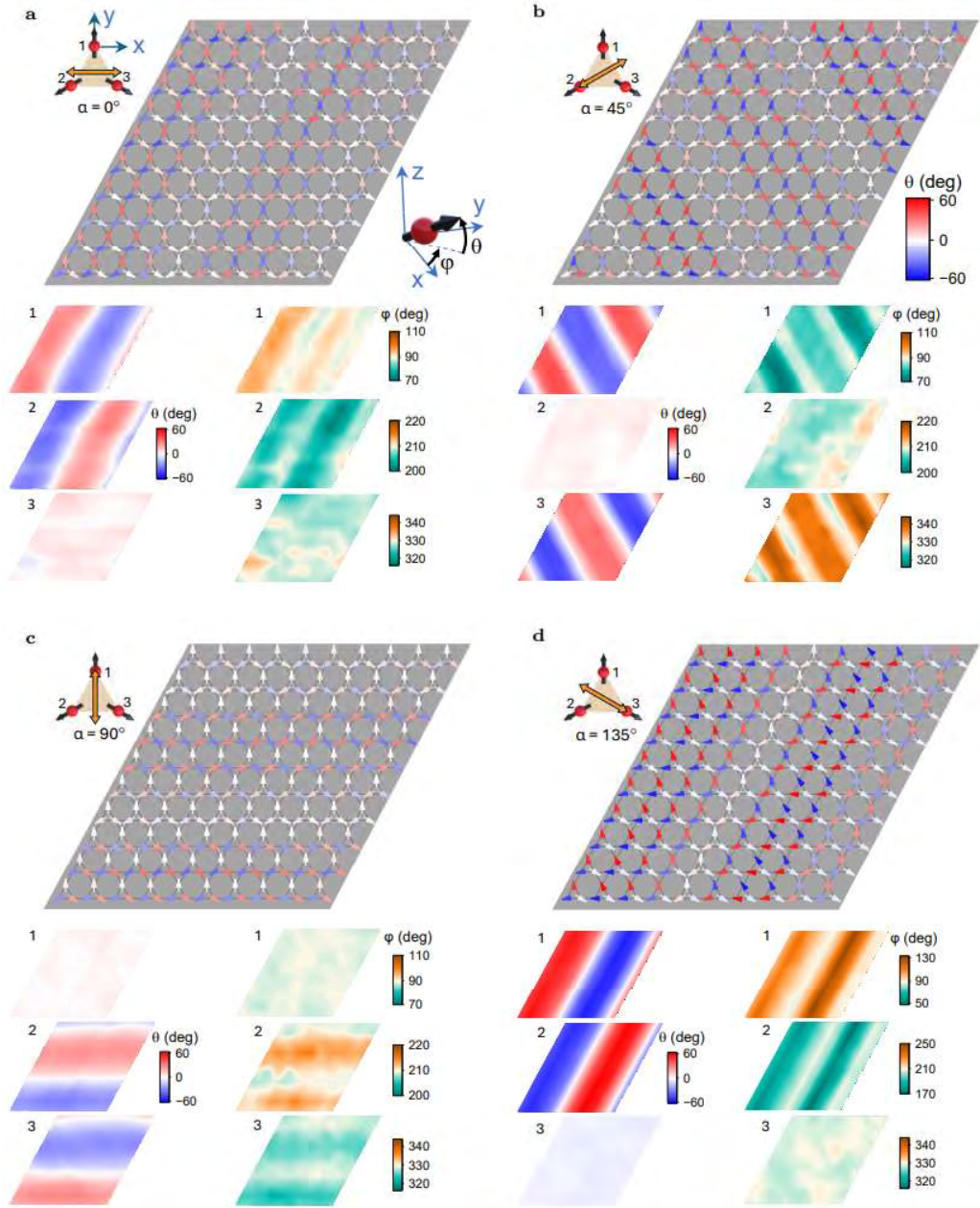

**Fig. S29**: **Spin spiral for $\Gamma^{4g}$.** As in **Fig. 5b,c**, but in the contour graphs, both the OOP spin tilt $\theta$ and IP spin tilt $\varphi$ are shown.

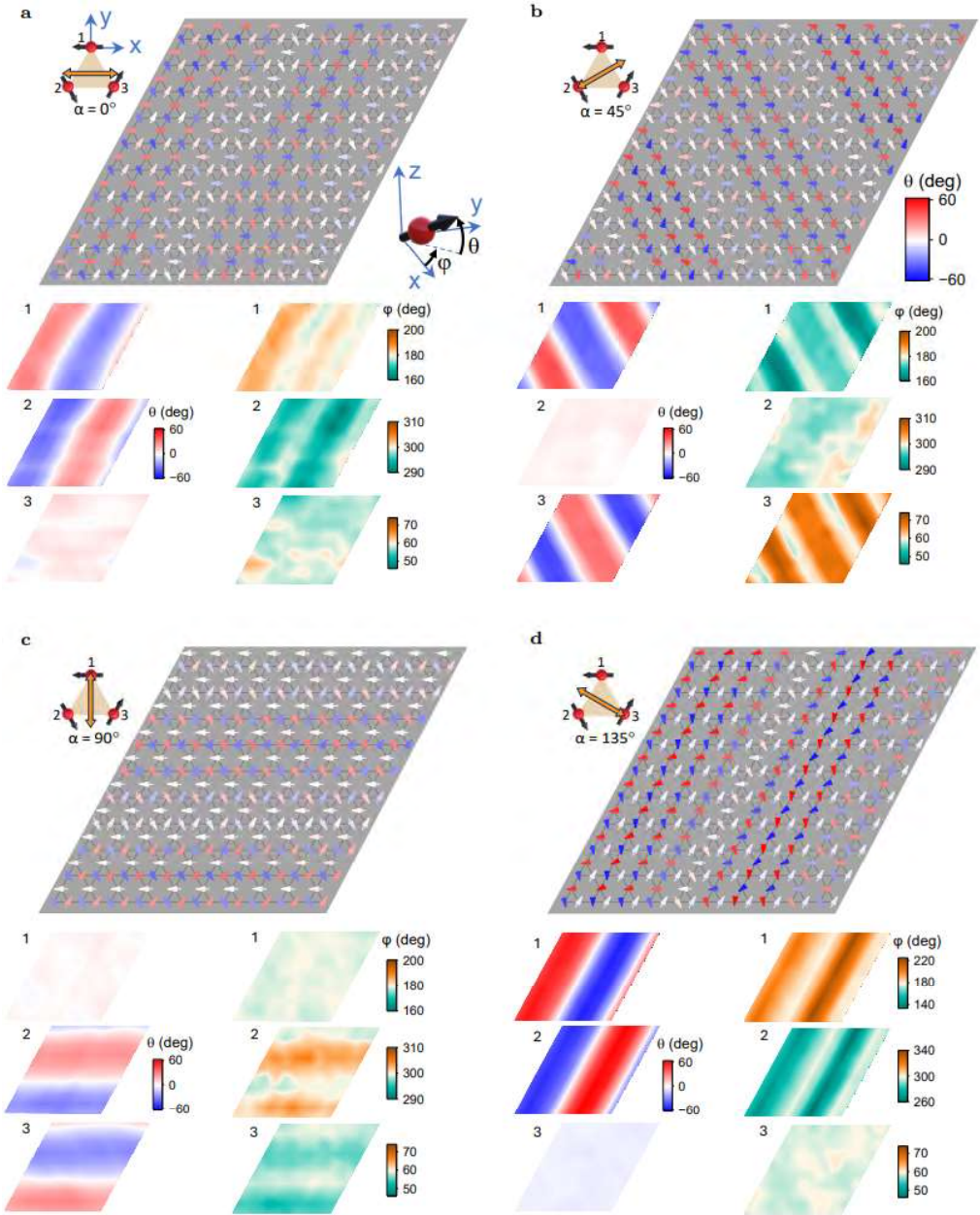

**Fig. S30**: **Spin spiral for $\Gamma^{5g}$.** As in **Fig. S29**, but magnetic configuration is $\Gamma^{5g}$.

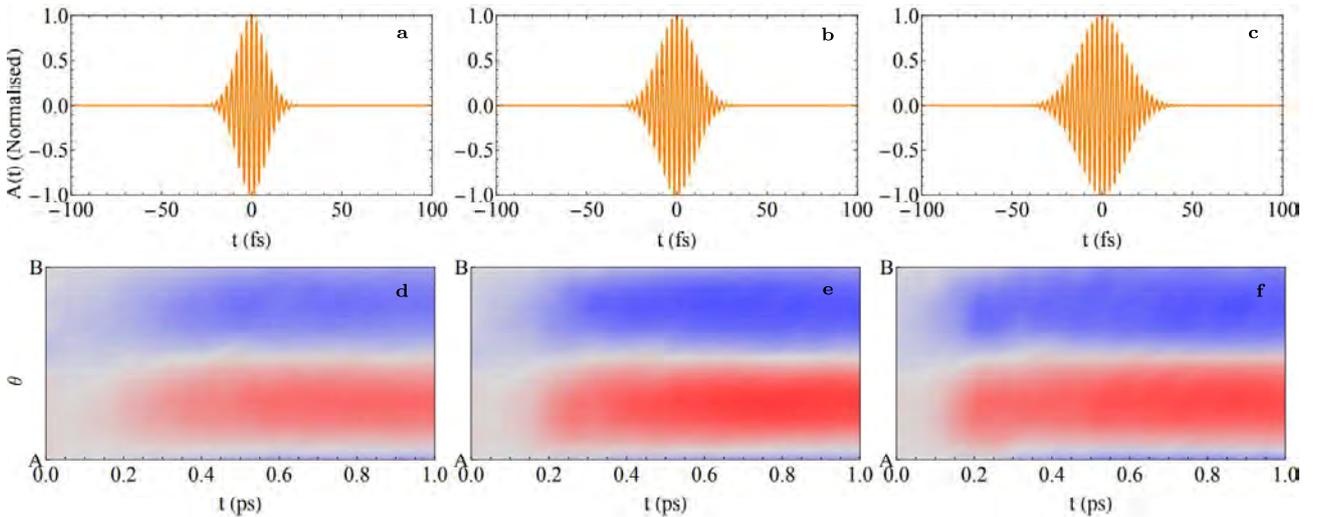

**Fig. S31**: **Different pulse durations. a, b, c** show temporal profile of exciting pulses with durations $s = 8$ fs, $s = 10$ fs and $s = 13$ fs, respectively. **d, e, f**, As in Fig. S27f, but the dynamics was excited with pulses from **a, b, c**, respectively.